\documentclass[11pt,a4paper]{article}                           

\usepackage{a4}          

\usepackage[UKenglish]{babel}
\usepackage{amsmath}
\usepackage{amssymb}
\usepackage[usenames]{color} 
\usepackage{multirow}
\usepackage{graphicx}
\usepackage{epstopdf}
\usepackage{array}
\usepackage{hhline}
\usepackage{cite}
\usepackage{microtype}
\usepackage[bf]{caption}
\usepackage{color}
\usepackage[usenames,dvipsnames]{xcolor}
\usepackage{subcaption}
\usepackage{pstool}

\usepackage{ifpdf}
\ifpdf
  \usepackage[pdftex,
    pdftitle={Backreacted Axion Field Ranges in String Theory},
    pdfauthor={},
    pdfsubject={},
    bookmarksopen, bookmarksnumbered, bookmarksopenlevel=2]{hyperref}
\fi
\def\hybrid{\topmargin -20pt    \oddsidemargin 0pt
        \headheight 0pt \headsep 0pt
        \textwidth 6.25in       
        \textheight 9 in       
        \marginparwidth .875in
        \parskip 5pt plus 1pt 
          \jot = 1.5ex
   }
\hybrid
\numberwithin{equation}{section}
\numberwithin{table}{section}

\newcommand{\beq}{\begin{equation}}  \newcommand{\eeq}{\end{equation}}
\newcommand{\bal}{\begin{aligned}}   \newcommand{\eal}{\end{aligned}}
\newcommand{\bea}{\begin{eqnarray}}  \newcommand{\eea}{\end{eqnarray}}

\newcommand{\bmat}{\left(\begin{array}}
\newcommand{\emat}{\end{array}\right)}
\def\om{\omega}

\newcommand{\nn}{\nonumber}

\newcommand{\I}{\text{Im}}
\newcommand{\R}{\text{Re}}

\usepackage{cancel}

\newcommand{\be}{\begin{equation}}
\newcommand{\ee}{\end{equation}}

\newcommand{\half}{\frac{1}{2}}

\begin{document}

\baselineskip=14pt
\parskip 5pt plus 1pt

\vspace*{-1.5cm}
\begin{flushright}    
  {\small
  }
\end{flushright}

\vspace{2cm}
\begin{center}        
  {\LARGE Backreacted Axion Field Ranges in String Theory}
\end{center}

\vspace{0.75cm}
\begin{center}        
Florent Baume, Eran Palti
\end{center}

\vspace{0.15cm}
\begin{center}        
  \emph{Institut f\"ur Theoretische Physik, Ruprecht-Karls-Universit\"at, \\
             Heidelberg, Germany}
             \\[0.15cm]
 
\end{center}

\vspace{2cm}


\begin{abstract}
String theory axions are interesting candidates for fields whose potential might be controllable over super-Planckian field ranges and therefore as possible candidates for inflatons in large field inflation. Axion monodromy scenarios are setups where the axion shift symmetry is broken by some effect such that the axion can traverse a large number of periods potentially leading to super-Planckian excursions. We study such scenarios in type IIA string theory where the axion shift symmetry is broken by background fluxes. In particular we calculate the backreaction of the energy density induced by the axion vacuum expectation value on its own field space metric. We find universal behaviour for all the compactifications studied where up to a certain critical axion value there is only a small backreaction effect. Beyond the critical value the backreaction is strong and implies that the proper field distance as measured by the backreacted metric increases at best logarithmically with the axion vev, thereby placing strong limitations on extending the field distance any further. The critical axion value can be made arbitrarily large by the choice of fluxes. However the backreaction of these fluxes on the axion field space metric ensures a precise cancellation such that the proper field distance up to the critical axion value is flux independent and remains sub-Planckian. We also study an axion alignment scenario for type IIA compactifications on a twisted torus with four fundamental axions mixing to leave an axion with an effective decay constant which is flux dependent. There is a choice of fluxes for which the alignment parameter controlling the effective decay constant is unconstrained by tadpoles and can in principle lead to an arbitrarily large effective decay constant. However we show that these fluxes backreact on the fundamental decay constants so as to precisely cancel any enhancement leaving a sub-Planckian effective decay constant.
\end{abstract}

\thispagestyle{empty}
\clearpage

\setcounter{page}{1}


\newpage

\tableofcontents

\section{Introduction}

From the perspective of Quantum Field Theory (QFT) large field inflation has a number of advantages over its small field counterpart: it can be realised by extremely simple potentials and requires no fine-tuning of an inflection point or the initial value of the field. However, as yet, no significant primordial tensor modes have been observed \cite{Ade:2015tva}. One possible explanation could be that the QFT perspective is misleading because there are Quantum Gravity (QG) constraints which actually obstruct this naively natural possibility, for example by limiting the available field range over which the potential can remain flat enough to be sub-Planckian. The question of the possible field excursion distances in an ultraviolet (UV) theory of gravity is one which has bearing on our understanding of current and future experimental results, through the connection to primordial tensor modes, while on the other hand requires a well-understood quantum theory of gravity to answer with significant detail and confidence. 

A possible objection to large field inflation, or more generally to super-Planckian field displacements, from an effective QFT perspective is that one might expect an infinite number of operators controlled by powers of the field over the Planck scale to appear once UV gravitational physics has been integrated out. Such a situation would imply a breakdown of the effective theory for super-Planckian field values. An effective QFT answer to this could be that there is an additional symmetry in the theory which is respected by the UV physics and which controls these operators beyond their naive dimensional analysis. In turn, the objection to this is that QG is expected to not respect any global symmetries which could control operators in the theory. At this point it is not clear how to proceed through general reasoning. Without knowing the UV theory we can not analyse the actual corrections from the UV physics, whether there is some symmetry that could control them and to what extent. One way to progress is to utilise general expectations of properties of QG, in particular along the lines of the Weak Gravity Conjecture (WGC), to restrict certain scenarios \cite{ArkaniHamed:2006dz,Rudelius:2015xta,Montero:2015ofa,Brown:2015iha,Bachlechner:2015qja,Brown:2015lia,Heidenreich:2015wga,Heidenreich:2015nta,Ibanez:2015fcv,Hebecker:2015zss}. Other approaches include an analysis of entropy bounds \cite{Conlon:2012tz,Kaloper:2015jcz}. In this paper we will adopt the approach of considering string theory as a UV completion and explicitly study these questions in such a framework. 

Within the string theory context we are interested in closed-string fields as candidates for fields that can support super-Planckian excursions and therefore possibly large field inflation. Such fields arising from a string compactification split into two classes, there are moduli and there are axions. The two differ in a number of ways, perhaps the most important one being that axions do not appear perturbatively (as an expansion in the moduli vevs) in the Kahler potential due to a perturbative shift symmetry. Moduli fields have been considered as possible candidates for fields that could support super-Planckian excursions, see \cite{Baumann:2014nda} for a review, \cite{Cicoli:2008gp} for early work and \cite{Broy:2015zba} for the most recent. However in doing so one faces two problems. The first is the generic one of controlling Planck scale suppressed operators since there is no UV symmetry to protect the moduli. The second problem is that the moduli values are also the parameters which control the effective theory such that their variation is bounded within a controlled effective theory. For too small moduli values there are large corrections to the effective theory, while for too large values the cut-off scale of the theory, for example the string or Kaluza-Klein scale, becomes low. The problem is amplified by the fact that closed-string moduli appear in the Kahler potential, and therefore in their own field space metric, in such a way that the canonically normalised fields are logarithmic in the moduli values. Therefore the displacement distance in the canonical field is exponentiated as a modulus variation which means one is fighting for control against an exponential. This may be possible if the coefficient in the exponent which multiplies the modulus could be controlled in some way, but it typically is only dependent on intersection numbers of the Calabi-Yau (CY) and such quantities rather than say a flux parameter. It is therefore difficult to ensure full control over super-Planckian field excursion in such a scenario, at least while keeping the cutoff scale of the theory sufficiently high, although we are not aware of a general proof against such a possibility. 

Closed-string axions do not appear to suffer from these difficulties at first glance. Their vacuum expectation value (vev) is not a control parameter for the effective theory and since they do not appear in the perturbative Kahler potential their field space metric is also independent of their vev. Further they are protected by a perturbative shift symmetry which is broken to a non-perturbative discrete symmetry. These properties make them attractive candidates for fields that could support super-Planckian displacements within a fully controlled setting. On the other hand an analysis of possible axion periodicity lengths appears to suggest that a super-Planckian period for a fundamental string axion is not achievable, see \cite{Banks:2003sx} for the key original paper and \cite{Conlon:2016aea} for the most recent analysis (while for example \cite{Kenton:2014gma,Kooner:2015rza} study possible exceptions). One possibility for avoiding this limitation is to consider mixing two or more fundamental axions to get an effective super-Planckian axion decay constant for an aligned combination \cite{Kim:2004rp}. This possibility has been studied in a string theory setting, see \cite{Baumann:2014nda} for a review and \cite{Long:2014dta,Ben-Dayan:2014lca,Hebecker:2015rya,Ruehle:2015afa,Gao:2014uha,Palti:2015xra,Kappl:2015esy} for recent work. Most relevant for the work in this paper is the analysis in \cite{Palti:2015xra} of type IIA compactifications. There it was shown that backreaction of the axion alignment parameter modified the metric on the axion field space so as to precisely cancel any enhancement of the axion periodicity. It was pointed out in \cite{Palti:2015xra} that in the case of more than two fundamental axions mixing, with an additional contribution to the superpotential to fix the additional combination,  it is possible that this cancellation of the enhancement may be modified. In this work we will study such a more general possibility by considering compactifications of type IIA string theory on a twisted torus rather than a CY setting. 

The primary focus of this work will be on a different idea for realising super-Planckian displacements with axions which is termed axion monodromy \cite{Silverstein:2008sg,McAllister:2008hb}. The basic idea behind it is to include effects in the vacuum which break the axion periodicity, for example a mass term, allowing it to traverse distances larger than a single period. For closed-string axions the effects which break the axion monodromy can be taken as branes, in IIA 6-branes as in \cite{Escobar:2015fda}, within IIB 5-branes as in \cite{McAllister:2008hb } and 7-branes as in \cite{Palti:2014kza}, or as background fluxes. The fact that a background flux in string theory can induce a non-periodic potential for an otherwise periodic axion is a well known effect in flux compactifications, though it has only recently been applied in the context of large field inflation \cite{Marchesano:2014mla,Blumenhagen:2014gta,Hebecker:2014eua,Arends:2014qca,Ibanez:2014kia,Blumenhagen:2015qda,Blumenhagen:2015kja,Blumenhagen:2014nba,Hayashi:2014aua,Hebecker:2014kva,Buchmuller:2015oma,Blumenhagen:2015xpa,Andriot:2015aza}. Recent work has also studied axion monodromy induced by non-flux effects \cite{Escobar:2015fda,Hebecker:2015tzo}. 

Both axion alignment and axion monodromy share the feature that they have some integer parameter $N$ such that for $N=1$ the field remains sub-Planckian and is then parametrically increased to super-Planckian values by dialing $N$ large. The crucial effect which we focus on in this paper is the backreaction of the parameter $N$ on the metric of the axion field space as it is dialed large. We focus on this as it is physics which has the right properties to form part of a censurship mechanism for large field displacements in string theory. The backreaction is gravitational physics and it interacts with the metric on the axion field space which determines the distance that the field travels. Further, while in QFT we are free to pick the field space metric as we like, typically it is considered to be a constant, in string theory it is a highly restricted and structured quantity. Ranging from coarse structures such as the universal logarithmic behaviour in the moduli of the Kahler potential signifying gravitational physics to fine details such as the structure of singularities on the field space which capture highly non-trivial non-perturbative quantum gravitational physics. Indeed the cancellation of N in the alignment scenarios found in \cite{Palti:2015xra} are examples of this general idea, and the form of the Kahler potential played a central role. If we let ourselves take the metric on the field space to be constant, or arbitrary, or neglect its dependence on $N$, this effect would not be captured.

In the case of axion monodromy the parameter $N$ is associated with the number of periods that the axion traverses. Its backreaction effect is therefore the backreaction of the energy density induced by the axion vev as it moves along its potential. For a sufficiently smooth and diluted flux background the backreaction of $N$ is captured at leading order by its effect on the other moduli values.\footnote{See \cite{Conlon:2011qp} for an analysis of backreaction in a non-dilute region within the context of large field inflation.} We would like to study this effect and determine its influence on the possible axion excursion distance. Since we must be careful to account for any QG features of string theory that could affect this calculation we would like to use a framework which is as clean and simple as possible. We will therefore study this effect within the moduli stabilisation framework of type IIA flux compactifications on a CY manifold \cite{Grimm:2004ua,DeWolfe:2005uu} or a twisted torus \cite{Villadoro:2005cu,Camara:2005dc} (see \cite{Grana:2005jc} for a more complete reference list on these topics). In such a setting moduli stabilisation is a completely perturbative process which means we only need to utilise tree-level expressions for the Kahler potential and superpotential. Further it is simple enough to allow us to solve for the backreaction of $N$ explicitly and precisely. Finally it is a setting for which the uplift to 10-dimensional supergravity solutions is well understood adding a further level of control. In this paper we will focus on the backreaction of the axion vev on its field space metric and thereby its excursion distance. There are a number of studies in the literature of the backreaction of axion vevs in axion monodromy scenarios within different settings which instead focus on the effect of flattening its potential or on the possible destabilisation of the vacuum. See for example \cite{Dong:2010in,Hebecker:2014kva,Buchmuller:2015oma,Bielleman:2015lka,Ibanez:2014swa,Dudas:2015lga,Dudas:2015mvk}. Perhaps most similar in a technical sense to our type IIA settings are the non-geometric compactifications studied in \cite{Blumenhagen:2015qda,Blumenhagen:2015kja,Blumenhagen:2014nba,Blumenhagen:2015xpa} which appear to share some features with our constructions. 

The paper is structured as follows. In section \ref{sec:axnmon} we study backreaction effects in axion monodromy scenarios, while in section \ref{sec:axionalign} we study backreaction effects in an axion alignment scenario. We summarise our results in section \ref{sec:summary}.

\section{Axion monodromy and backreaction}
\label{sec:axnmon}

In this section we analyse models of flux-induced axion monodromy in type IIA string theory. There are two types of axions in compactifications of type IIA string theory on Calabi-Yau orientifolds. The Ramond-Ramond (RR) axions which pair up with the complex-structure moduli and the dilaton, and the Neveu-Schwarz (NS) axions which pair up with the Kahler moduli. Turning on RR fluxes induces a potential for the NS axions and vice-versa. They then become so-called monodromy axions, and the parameter $N$ amounts to the number of times they traverse their period and so is measured by their vacuum expectation value. Therefore accounting for the effect of $N$ on the field space metric amounts to calculating the backreaction on the moduli of the axion vev.

Before entering into the calculation it is worthwhile to expand on the methodology we will utilise to study the backreaction and field range. If there is some constraint on field distances in string theory it is unclear what form it takes. The notion of a bound on a single field variation, tracing a line in scalar field space, is not well defined in the presence of multiple fields since a higher dimensional field space trivially supports an infinite length line. It may be that it is possible to formulate a constraint on the volume of field spaces, but in the absence of such a well defined formulation we adopt a different approach. Once we introduce a potential to the theory we can consider the dimensionality of the vacuum space of the theory. At the non-perturvative level one expects all fields to gain a potential which means that if the vacuum space is not empty then it is a collection of points. We can also consider the perturbative vacuum space which can be higher dimensional but still typically reduced from the dimensionality of the total field space. Then within this framework there are two natural ways to identify a one-dimensional field space along which we can test bounds on excursion distances. The first is the lightest direction in field space. This is the identification we will use when studying axion alignment. It is possible to think of this as a theory where we integrate out all the heavier fields to obtain an effective field theory for a single field and see if our knowledge of the UV origin of this theory constrains the effective field range. 

The second possibility for identifying a single-field space is to consider dropping the requirement for one field to lie in the vacuum. So we displace by hand a single field from the vacuum and ask how far can this field go. This is the approach we will adopt for studying axion monodromy models, with the field away from the vacuum being the massive axion. This is quite a different setting to the previous one and we should not think of integrating out the other fields to obtain some effective theory for the field we are displacing. Rather they must be kept in the theory and we must adjust their values appropriately to track their minima as a function of the displaced field, this is the backreaction effect which we crucially want to capture. If the other fields are much more massive than the field which is displaced from the vacuum then they will not change much over the displacement, but this does not have to be imposed for the procedure to be well-defined. Note that it could be that after displacing the field far enough away from a stable vacuum we reach a point where the minima for the other fields disappear and they become unstable themselves. In this case the whole theory would undergo some phase transition to a different vacuum. We will show examples of such cases but they will not be of central interest for us, we will focus on the cases where the other fields have well defined, though continuously changing, minima.

Once we identify a line path in field space we are interested in how far along this path the field can displace. We must consider the proper path length in field space, so for a space of fields spanned by $\nu^i$ coordinates with metric $g_{ij}$ we want to calculate 
\be
\Delta \phi \equiv \int_{\gamma}  \sqrt{g_{ij} \frac{\partial \nu^i}{\partial \rho} \frac{\partial \nu^j}{\partial \rho}  } d \rho \;. \label{defphpr}
\ee
Here $\rho$ is the world-line element along the path defined by $\gamma$ which is specified as an embedding $\gamma \;:\; \rho \rightarrow \nu^i \left(\rho\right)$. The important effect to capture is that $g_{ij}$ is a function of $\rho$ once the gravitational backreaction is accounted for, and the integral is then appropriately modified. Applying this to the type IIA string theory setting we will be  interested in the case where the path in (axion) field space is along a certain linear combination of fields
\be
\rho = \sum_i h_i \nu^i \;.
\ee
We can then simplify (\ref{defphpr}) to
\be
\Delta \phi = \int_{\rho_i}^{\rho_f} \left(h_i g^{ij} h_j \right)^{-\half}  d \rho \;. \label{defphprsim}
\ee
This is seen to be the integration of the canonical normalisation factor for the field $\rho$ from its initial values $\rho_i$ to its final one $\rho_f$.\footnote{Note that we work in conventions where a real scalar field has canonical kinetic terms $\left(\partial \phi\right)^2$, with no prefactor of $\half$.}

\subsection{Ramond-Ramond axions}
\label{sec:RRaxion}

We begin by analysing the displacement of a massive combination of RR axions from the minima. We will initially study a simplified model where each sector of moduli, complex-structure, Kahler, and dilaton have only one representative variation. This captures the behaviour of the system under a universal scaling of the moduli, and can be thought of as restricting the moduli values to be equal. We will see that the important physics manifests already in this simple setting. In section \ref{sec:RRaxionCY} we will then generalise the results to a realistic Calabi-Yau system of moduli in the RR axions sector (which is the complex-structure moduli sector). In section \ref{sec:RRaxiontt} we will further generalise the setting to the case of a twisted torus.

\subsubsection{Single field models}
\label{sec:RRaxionsingle}

The starting point is the model studied in \cite{Palti:2015xra} which considered a simplified version of the IIA on CY model 
\begin{equation}\label{simsup}
	K = - \log s - 3\log u - 3\log t \;, \;\;\; W = e_0 + i h_0 S - i h_1 U + \frac{i}{6} m T^3 \;. 
\end{equation}
Here the $h_i$ are NS fluxes, while $e_0$ and $m$ are RR fluxes. Note that we use the notation of \cite{Camara:2005dc}. The superfields are
\begin{equation}
	S= s+i\sigma \;,\; U= u+ i \nu\;,\; T = t+ iv \;. \label{sfexpa}
\end{equation}
There are two simplifications which enter this construction. The first is that, as discussed above, we have taken only a single modulus in the Kahler and complex-structure moduli sector. The second simplification is that we have turned off some fluxes, see (\ref{simgen}) for the full set of fluxes. There is one axion combination $\rho$ which becomes massive due to the fluxes and therefore suffers a monodromy effect
\begin{equation}
	\rho = e_0-h_0\sigma + h_1 \nu \;. \label{defrho}
\end{equation}
As discussed above, to capture the backreaction effect of the axion vev we study the values of the moduli which solve the equations
\begin{equation}\label{minpot}
	\partial_{T}V=\partial_{u}V=\partial_s V = 0 \;,
\end{equation}
as a function of the vev of $\rho$. Here $V$ is the scalar potential of the theory following from (\ref{simsup}). We do not impose the minima equations for the two axion combinations of $\sigma$ and $\nu$, this is because one combination is perturbatively massless while the other we would like to displace from its minimum. 

The solution to these equations was presented in \cite{Palti:2015xra} and reads 
\begin{equation}\label{amsolsim}
	s = \alpha\frac{\rho}{h_0}\;,\;\; u = -3 \alpha \frac{\rho}{h_1} \;,\;\; t = 1.96 \left(\frac{\rho}{m} \right)^{\frac13}\;,\;\; v=0\;, 
\end{equation}
with $\alpha\simeq 0.38$. Note that this solution is only valid for sufficiently large values of $\rho$. In fact the solution does not flow to a physical minimum for any value of $\rho$ which can be seen by noting that there is no physical superymmetric vacuum for the system (due to the restriction on the superpotential). Nonetheless it is a useful model for capturing the axion back-reaction, indeed the crucial point is that we find for the axion field space metric $\sqrt{g_{\rho \rho}} \sim s^{-1} \sim \rho^{-1}$. This means that the metric on the field space of the axion is modified so that the canonically normalised field distance is only logarithmic in the vev of $\rho$ \cite{Palti:2015xra}. More precisely we obtain from (\ref{defphprsim}) 
\be
\Delta \phi = \int_{\rho_i}^{\rho_f} \frac12 \left(\left(h_0 s\right)^2+ \frac13 \left(h_1 u\right)^2 \right)^{-\frac12} d\rho \simeq 0.7 \log \left(\frac{\rho_f}{\rho_i}\right) \;. \label{phicsimpl}
\ee
There are two qualitative features of (\ref{phicsimpl}) to highlight. The most important feature is that the proper field distance is only logarithmic in the axion variation. This is the type of behaviour we expect from moduli fields and not axions. In this sense, once backreaction is accounted for, inducing a potential for an axion makes it behave like a modulus. The second important features is that the prefactor $0.7$ is flux independent. Therefore there is no parameter to adjust in the model to make the field excursion large while keeping the logarithmic term small. While formally the field distance is unbounded, as discussed in the introduction it is not likely to be possible to obtain super-Planckian displacements in such a setting. Indeed it is precisely this logarithmic behaviour of moduli which we attempted to avoid by working with axions since an exponentially large variation of the moduli is difficult to support in an effective theory which is under control. And from (\ref{amsolsim}) we see that indeed the moduli  scale exponentially with the proper axion field distance.

Let us consider the generality of the result of $\Delta \phi \sim \log \rho$. To do this it is useful to notice that the equations determining the values of the moduli (\ref{minpot}) satisfy a scaling symmetry
\begin{equation}\label{ssym}
\rho \rightarrow \lambda \rho \;,\;\; s \rightarrow \lambda s \;,\;\; u \rightarrow \lambda u \;,\;\; T \rightarrow \lambda^{\frac13} T \;. 
\end{equation}
When we solve for the moduli in terms of $\rho$ and the fluxes, only the $\rho$ field carries weight under this symmetry and therefore $s$ and $u$ must be proportional to it. This argument is sufficient to establish the behaviour (\ref{phicsimpl}) up to a constant of proportionality factor. Now consider a general CY compactification with an arbitrary number of moduli. The system still satisfies the symmetry (\ref{ssym}) with an equal scaling of all the moduli because of the logarithmic behaviour of the Kahler potential. Therefore there must be an overall proportionality of all the $u_i$ moduli to $\rho$ which is sufficient to establish the logarithmic behaviour of (\ref{phicsimpl}). The proportionality factor $0.7$ in the universal behaviour (\ref{phicsimpl}) however can in principle depend on dimensionless fluxes, such as the $h_i$, in the case of an arbitrary CY manifold. In section \ref{sec:RRaxionCY} we study this problem and show that it is flux independent and of order one. 

For the superpotential (\ref{simsup}) the only flux parameter which carries weight under the scaling symmetry (\ref{ssym}) is $e_0$. This is particularly simple since it can just be absorbed into the definition of $\rho$. If we consider the most general superpotential for type IIA on a CY there are additional fluxes which scale with the symmetry (\ref{ssym}). The simplified version takes the form
\be
 W = e_0 + i h_0 S - i h_1 U + i e_1 T - q T^2 + \frac{i}{6} m T^3 \;. \label{simgen}
\ee
The fluxes of type $e_1$ and $q$ carry weights under the symmetry (\ref{ssym}) of $\frac23$ and $\frac13$ respectively. Their introduction implies a number of important changes. First, as we will show, the theory now develops physical minima. Secondly they imply a modification to the general argument leading to the behaviour (\ref{phicsimpl}) which accounts for the dimensionful fluxes. The solutions for the moduli in terms of $\rho$ now fall into two classes: those which in the limit $\left\{e_1,q\right\} \rightarrow 0$ reduce to the previous solution, and those which break down in this limit. We will turn to studying the solutions in detail soon but first let us make some general statements about the first class, which for $\left\{e_1,q\right\} \rightarrow 0$ reduces to (\ref{phicsimpl}). These solutions must have some functional form for the moduli in terms of $\rho$ such that when $\rho$ is larger than some critical value $\rho_{\mathrm{crit}}$ set by the magnitude of the fluxes which break the symmetry $\rho_{\mathrm{crit}} \sim \left(e_1^\frac32+q^3\right)$ they converge to (\ref{amsolsim}). The first thing to observe is therefore that by taking large fluxes we can delay the onset of the scaling behaviour (\ref{amsolsim}) arbitrarily far in $\rho$ distance. The fluxes therefore act to shield the moduli from the axion vev backreaction. However, even though the variation in $\rho$ can be extended parametrically far through this method we can see that this will not imply an arbitrarily far proper field distance. The rough argument is that if we assume that the moduli which control the axion field space metric, $s$ and $u$, remains approximately constant up to $\rho \sim \rho_{\mathrm{crit}}$ then since at $\rho_{\mathrm{crit}}$ their values go like $s \sim \rho_{\mathrm{crit}}$ this will be their approximate value over the regime of small $\rho$. The proper axion field distance $\Delta \phi$ up to $\rho_{\mathrm{crit}}$ will therefore behave as
\be
\Delta \phi \sim \frac{\Delta \rho}{\left. \left\{s,u\right\} \right|_{\rho_{\mathrm{crit}}}} \sim \frac{\rho_{\mathrm{crit}}}{\rho_{\mathrm{crit}}} \sim 1 \;.
\ee
So while $\rho_{\mathrm{crit}}$ may be arbitrarily large, controlling the backreaction for arbitrarily large field distances, its value cancels in the proper field distance. The argument presented is quite general but imprecise and the rest of this section will be dedicated to essentially filling in the missing details in it. 

There will be two steps to improving the argument just presented for the structure of the proper length field variation. First we would like to make it more quantitative keeping track of the relevant coefficients, and then we will determine the actual values of these coefficients. 
It is useful to introduce some coordinate redefinitions. First we absorb some of the fluxes into the definition of the moduli and other fluxes
\be
\tilde{s} = h_0 s \;, \tilde{u} = -h_1 u \;, \tilde{T} = T m^{\frac13} \;, \tilde{e}_1 = e_1 m^{-\frac13} \;, \tilde{q} = q m^{-\frac23} \;, \label{1modresc}
\ee
and introduce the flux combination
\be
f \equiv - \tilde{e}_1 - 2\tilde{q}^2 \;. \label{fdef}
\ee
An important role in our analysis will be played by the particular moduli values which correspond to a physical supersymmetric vacuum of the system. In this vacuum the fields take the values 
\be
\tilde{s}_0 =\frac{\tilde{u}_0}{3} = \frac{\tilde{t}_0^3}{15}= \frac29 \sqrt{\frac{10}{3}} f^{\frac32} \;,\; \rho_0 = \frac23 \tilde{q} \left(3f +2 \tilde{q}^2 \right)  \;,\; \tilde{v}_0 = -2\tilde{q} \;, \label{susymod}
\ee
where we take $h_0>0$, $h_1<0$ and $m>0$. In analysing the potential it is convenient to shift the axion definitions by their supersymmetric minimum values
\be
\rho' \equiv \rho - \rho_0 \;,\;\; v' \equiv \tilde{v} - \tilde{v}_0\;. \label{rpvp}
\ee
This is useful because after this shift the scalar potential, up to an overall constant factor, only depends on the single flux parameter $f$ explicitly
\be
V = h_0 h_1^3 m \tilde{V}\left(\tilde{s},\tilde{u}_1,\tilde{t},v',\rho',f\right) \;. \label{1parpot}
\ee
Therefore solutions to the equations (\ref{minpot}) will depend on only one parameter in these variables. The dependence of the potential on only one flux parameter $f$ can be understood as follows. Three flux parameters in the superpotential can be absorbed into moduli rescaling as in (\ref{1modresc}) which leaves three parameters $e_0$, $e_1$ and $q$. The theory however also respects two shift symmetries, one for the RR axions and one for the NS axion which can be used to absorb two more flux parameters into a shift in the axions. This leaves one flux parameter, which through our shifts is $f$. Note that we could equally use the shift symmetries to set $e_0$ and $q$ to zero for example.

The quantity of interest for us is the proper distance traversed by the massive axion field $\rho$ up to its critical value, as given in (\ref{defphprsim})
\be
\Delta \phi = \int_{\rho_i}^{\rho_f} \left(h_i g^{ij} h_j \right)^{-\frac12} d\rho = \int_0^{\rho'_{\mathrm{crit}}} \frac12 \left(\tilde{s}^2+ \frac13 \tilde{u}^2 \right)^{-\frac12} d\rho' = g\left(\frac{\rho'_{\mathrm{crit}}}{f^{\frac32}}\right) = r \;, \label{delphigen}
\ee
where $g$ is some arbitrary function which depends only on the shown ratio of $\rho'_{\mathrm{crit}}$ and $f$, and $r$ is a flux-independent number. The important point is that the result is independent of any flux parameters. In this analysis we have taken with generality the initial value at $\rho'=0$. The non-trivial step is the third equality since $\tilde{s}$ and $\tilde{u}$ are some complicated functions of $\rho'$. However recall that we still have the symmetry (\ref{ssym}) of the system under which $f$ carries weight $\frac23$ and $\rho'$ carries weight 1. Since $\Delta \phi$ must be dimensionless under this symmetry it can only depend on the appropriate ratio of $\rho'_{\mathrm{crit}}$ and $f$. The final step uses the fact that $\rho'_{\mathrm{crit}}$ scales with $f^{\frac32}$ since it is only $f$ which breaks the symmetry that leads to the solution \ref{amsolsim}. The end result is some flux-independent number which is expect to be of order one, to which we now turn.

We proceed now to analyse the structure of the scalar potential in detail to determine the precise numerical factors in the previous analysis. Consider a solution to (\ref{minpot}), this will satisfy
\begin{equation}\label{constru}
-\frac34e^{-K}\left(3 \tilde{s} \frac{\partial \tilde{V}}{\partial \tilde{s}} - \tilde{u} \frac{\partial \tilde{V}}{\partial \tilde{u}}\right) = \left(3 \tilde{s} - \tilde{u} \right)\left( 6\tilde{s} - \tilde{t}^3 + 2 \tilde{u}\right) = 0 \;.
\end{equation}
Therefore turning points of the potential split into two branches 
\bea
\mathrm{Branch\;1}:\; \tilde{u} &=& 3 \tilde{s} \;, \nonumber \\
\mathrm{Branch\;2}:\; \tilde{u} &=& - 3 \tilde{s} + \frac12\tilde{t}^3  \;. \label{branches}
\eea
From (\ref{susymod}) we see that only the first branch supports a supersymmetric minimum, and only for $f>0$. However a turning point of the potential occurs for both branches, for either sign of $f$, at the point $\rho'=v'=0$. These are then non-supersymmetric minima in general. Analysing the Hessian at these turning points  shows that for the supersymmetric turning point of branch 1 there is one negative eigenvalue, while for the non-supersymmetric minima of branch 2 all the eignevalues are positive. In the case of a negative eigenvalue it lies above the Breitenlohner-Freedman (BF) bound and so all these turning points are stable minima. These minima will form the starting points for our axion excursions in $\rho'$. As we move away from the minimum in $\rho'$ the stability with respect to the other directions continues to hold. For clarity we will henceforth restrict to $f>0$ so that the minimum of branch 1 is supersymmetric, while the branch 2 minimum is non-supersymmetric.\footnote{We have analysed also the cases for $f<0$. We find similar behaviour with the only key difference being that for this sign of flux excursions for large $\rho'$ along branch 2 destabilise the potential such that the turning points in the moduli disappear and the theory then undergoes a phase transition to a new vacuum. While interesting this limits the excursion distances in field space and is not the focus of this work.} 

As we move $\rho$ away from its supersymmetric minimum the other moduli will adapt as in (\ref{minpot}). Consider branch 1 in (\ref{branches}).  The system is quite complicated but we could solve it numerically and match the result onto a function. We find that to good accuracy the following function matches the numerical analysis
\begin{equation}\label{sols}
	\tilde{s} = \left[\left(\alpha\rho'\right)^4 + \beta f^3 \left(\rho'\right)^2 + \tilde{s}_0^4\right]^{\frac{1}{4}} \;. 
\end{equation}
Here $\alpha$ is as in (\ref{amsolsim}), $\alpha \simeq 0.38$, and $\beta \simeq 0.05$.\footnote{For $f<0$ we find a similar fit but with $\beta \simeq 0.03$.} This shows the interpolating behaviour between the supersymmetric minimum value for $\rho'$ and the large vev limit (\ref{amsolsim}). We can therefore define $\rho'_{\mathrm{crit}}$ as the value of $\rho'$ for which the first term in (\ref{sols}) becomes equal in magnitude to the sum of the other two. This gives
\be
\rho'_{\mathrm{crit}} \simeq 1.7 f^{\frac32}\;.
\ee
We can now evaluate the general expression (\ref{delphigen}) using the solution, which gives 
\be
\Delta \phi \simeq 0.9 \;. \label{dpbr11m}
\ee
This gives the precise numerical evaluation of the general structure discussed previously. The key result is that the canonically normalised field distance is independent of any fluxes and is of order one. 

\begin{figure}[tbp]
	\centering
	\begin{subfigure}[b]{0.5\textwidth}
		\centering
		\includegraphics[width=\textwidth]{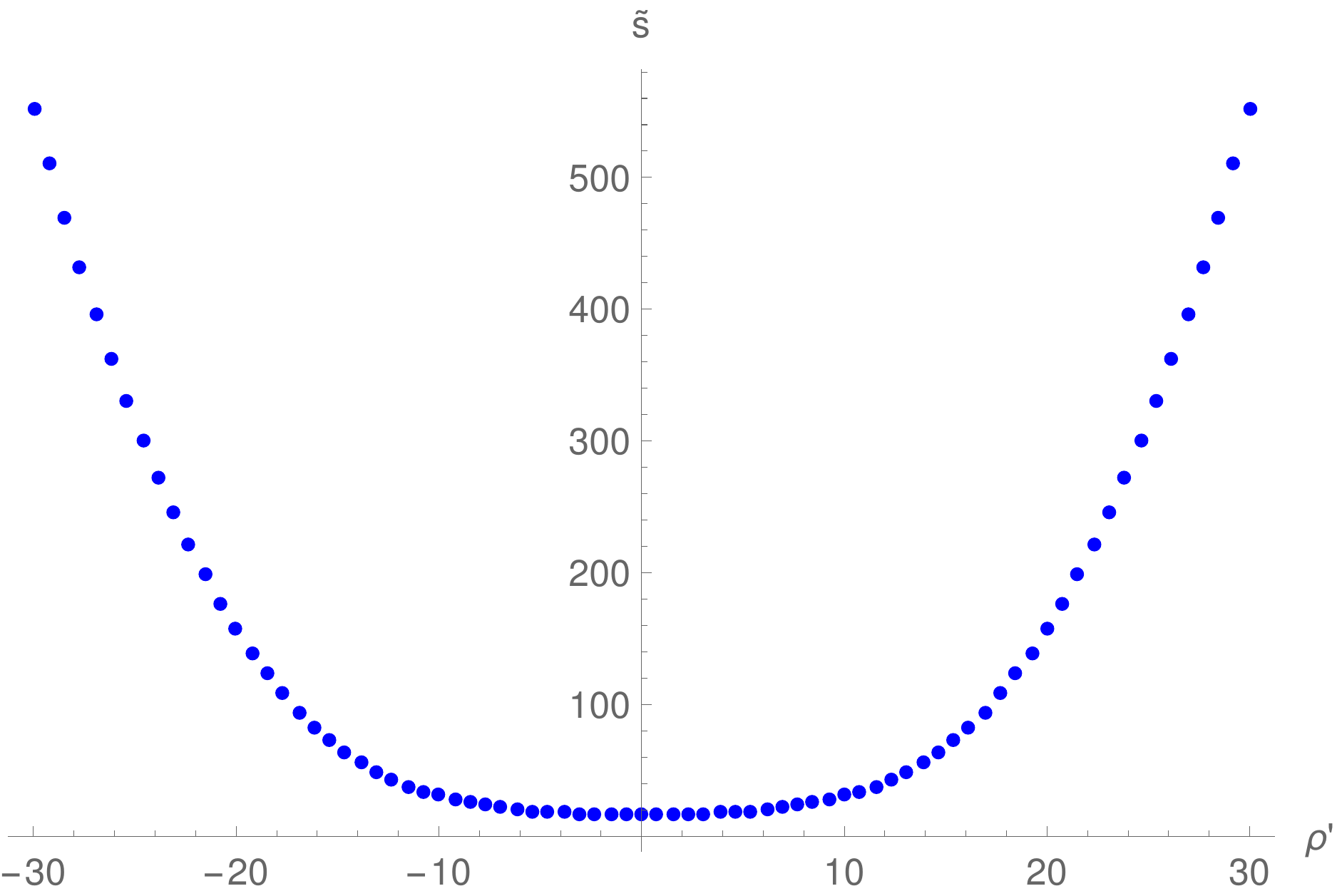}
	\end{subfigure}~
	\begin{subfigure}[b]{0.5\textwidth}
		\centering
		\includegraphics[width=\textwidth]{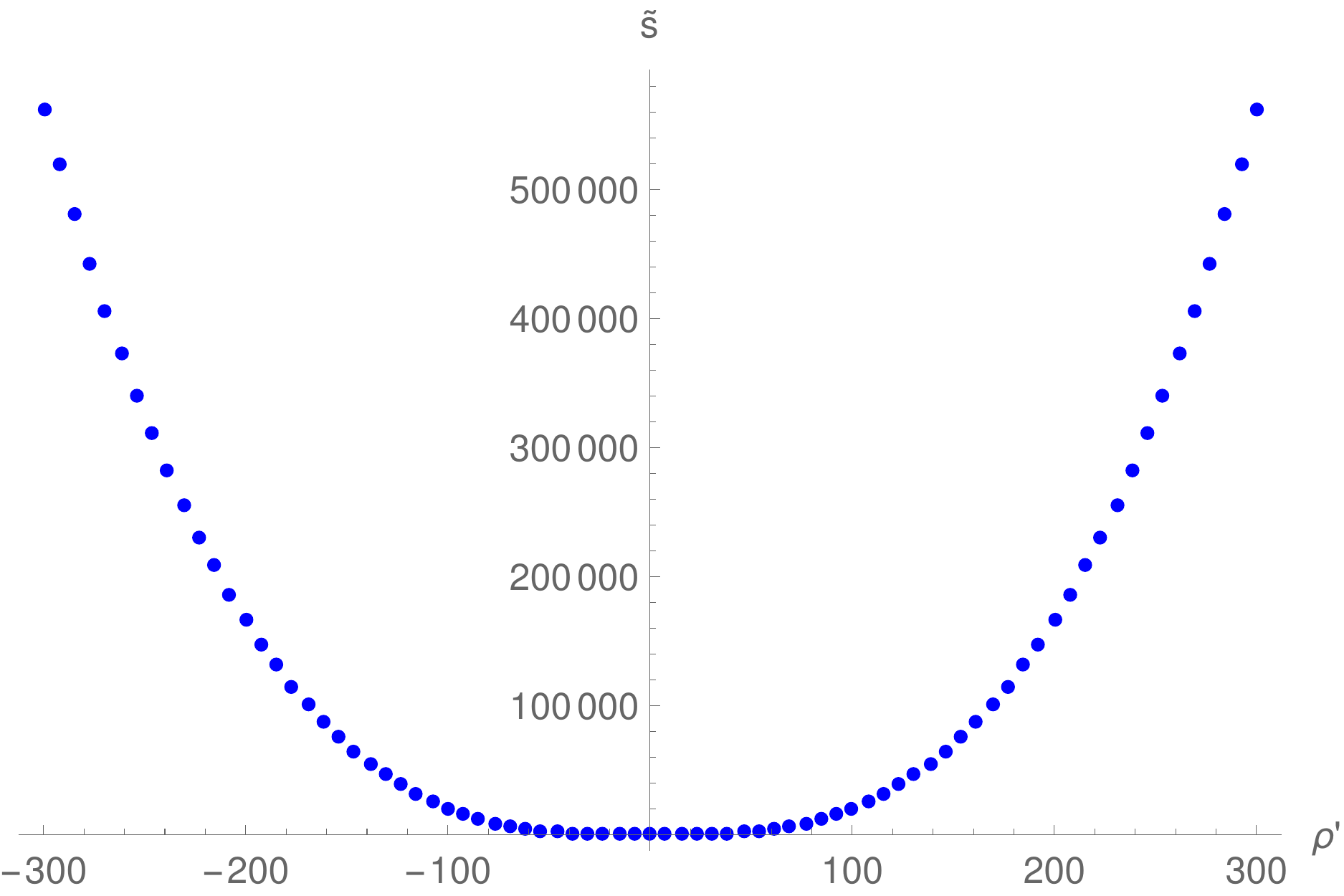}
	\end{subfigure}\\
	\begin{subfigure}[b]{0.5\textwidth}
		\centering
		\includegraphics[width=\textwidth]{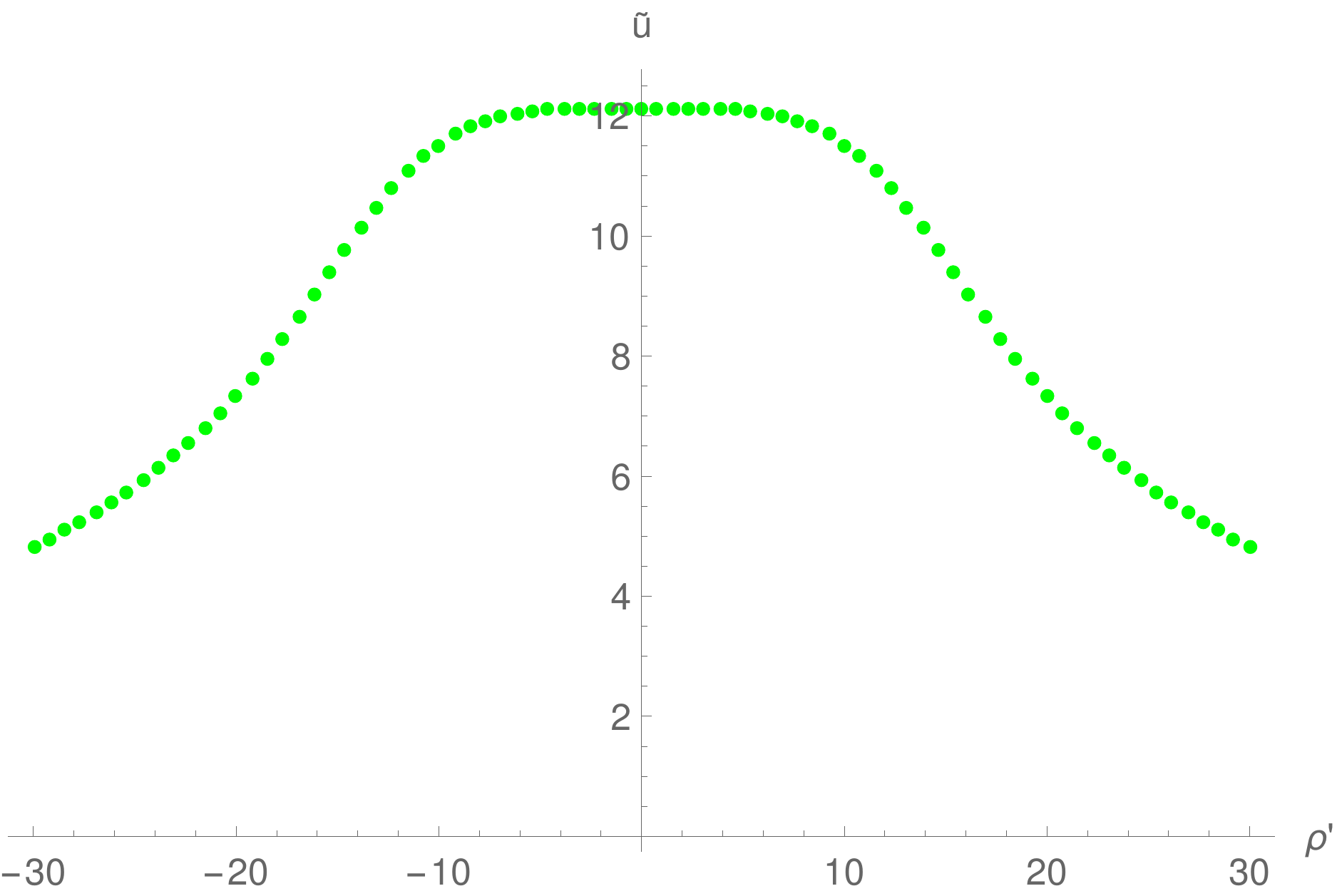}
	\end{subfigure}~
	\begin{subfigure}[b]{0.5\textwidth}
		\centering
		\includegraphics[width=\textwidth]{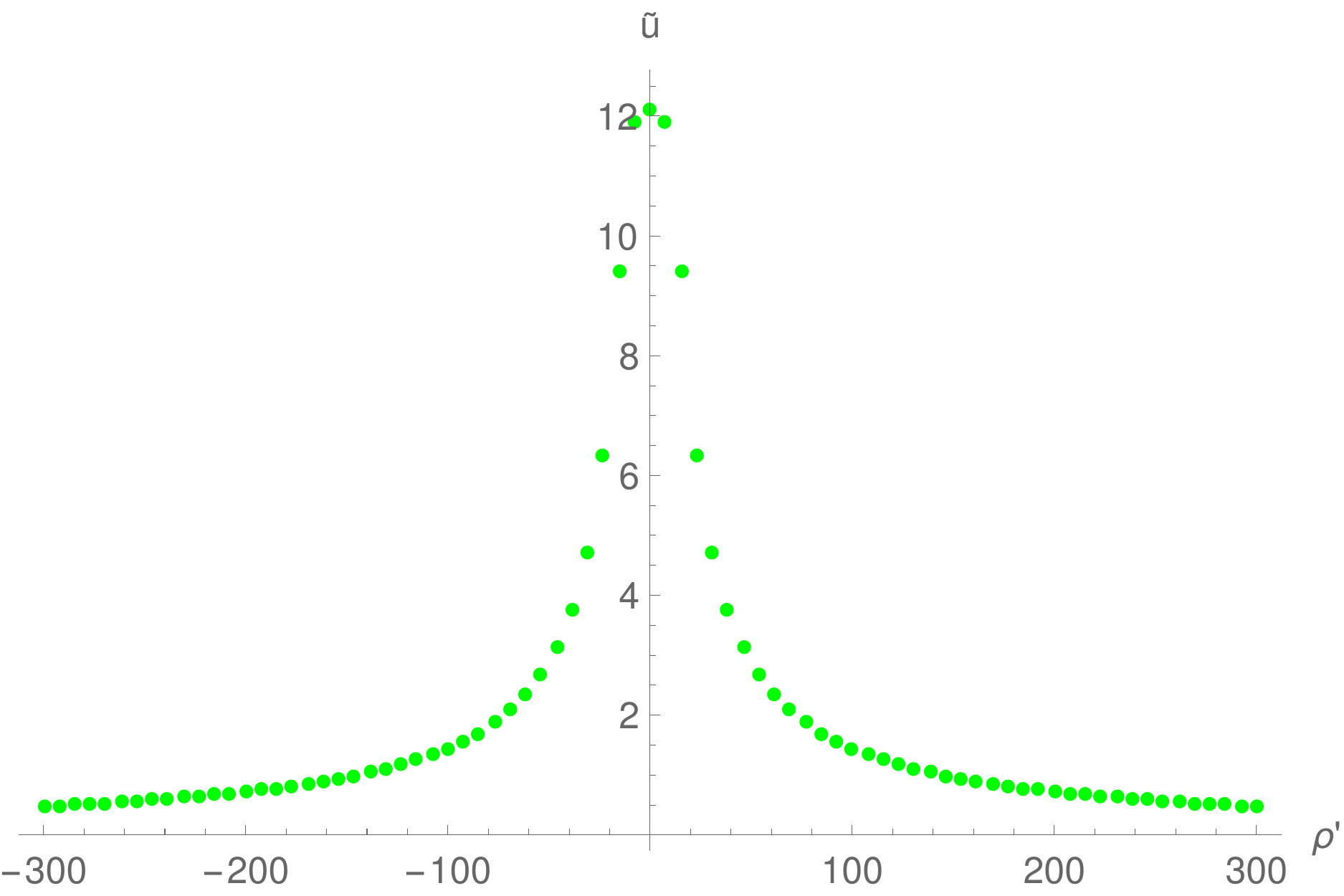}
	\end{subfigure}\\
	\caption{Plots showing the moduli $\tilde{s}$ and $\tilde{u}$ as a function of $\rho' = \rho - \rho_0$ for displacement of $\rho'$ along branch 2 of (\ref{branches}). The plots are for flux value $f=6$ and show the same function over two different ranges so as to show the behaviour up to $\rho_{\mathrm{crit}} \simeq f^{\frac32} \simeq 15$, and the asymptotic behaviour of $\tilde{u}$.}
	\label{fig:second branch}
\end{figure}

We performed a similar evaluation for the non-supersymmetric branch 2 in (\ref{branches}). We did not derive an analytic expression for the moduli as a function of $\rho'$ but studied it numerically. We find that for $\tilde{s}$ the large $\rho'$ scaling regime takes the form
\be
\tilde{s} \simeq 1.7 \left(\frac{\rho'}{f}\right)^3 \;.
\ee
The critical value of $\rho'$ where this regime begins is again around $f^{\frac32}$ as shown in figure \ref{fig:second branch}. The modulus $\tilde{u}$ instead asymptotes to zero but for values of $\rho' \gg \rho_{\mathrm{crit}}$ as shown in figure \ref{fig:second branch}. We can approximate to a decent accuracy the distance traveled by the canonical field up to $\rho'_{\mathrm{crit}}$ by taking $\tilde{s}$ and $\tilde{u}$ to be constants over that distance which gives then $\Delta \phi \simeq 0.5$. Note that because $\tilde{s}$ scales with a larger power of $\rho'$ after the critical point than for branch 1 the distance up to $\rho' \rightarrow \infty$ is not even logarithmically divergent but is finite. It gets cutoff very quickly giving an increase in $\Delta \phi$ of order a percent.

We have found that although the strong axion backreaction can be delayed arbitrarily far in terms of the axion vev, the proper field distance is flux independent. The cancellation of the flux parameters in the result for the proper length in field space is strikingly similar to a flux cancellation found in the case of axion alignment in \cite{Palti:2015xra}. Indeed the two can be related naturally. We can approximate the moduli to be independent of $\rho'$ for values $\rho' < \rho'_{\mathrm{crit}}$ and write the distance that the axion traverses before strong backreaction as $\Delta \phi \sim \rho'_{\mathrm{crit}} f_{\rho'} \sim N f_{\rho'}$ where $N=f^{\frac32}$ in terms of the flux. Here $f_{\rho'}$ is the normalisation factor appearing in the proper field distance (\ref{defphprsim}) evaluated at $\rho'< \rho'_{\mathrm{crit}}$, which can be defined as the axion decay constant for $\rho'$. Now consider an axion alignment scenario between two axions, labeled by $\nu_1$ and $\nu_2$. The effective massless axion combination $\psi$ appears in two instantons, in one of them with an effective decay constant which, before accounting for back-reaction effects, is enhanced by $N$ so that $f^{1}_{\psi} \sim N f_{\nu_2}$  and in the other one with an effective decay constant which is not enhanced $f^{2}_{\psi} \sim f_{\nu_2}$. Here $f_{\nu_i}$ are the fundamental decay constants for the $\nu_i$ axions. After accounting for backreaction, as in \cite{Palti:2015xra}, one finds that $f_{\nu_2} \sim \frac{f_{\nu_1}}{N}$ so that $f^{1}_{\psi}$ does not enhance. This scaling with the parameter $N$ implies an identification of the naively enhanced axion decay constant in the alignment scenario with the proper field length in monodromy $f^1_{\psi} \leftrightarrow \Delta \phi$, and the un-enhanced decay constant of the second instanton in alignment with the monodromy axion decay constant $f^2_{\psi} \leftrightarrow f_{\rho'}$. This map can be thought of as zooming into the origin of the large effective axion period in an axion alignment scenario where the potential looks approximately quadratic rather than sinusoidal and therefore is similar to an axion monodromy scenario. On top of the quadratic potential there is an oscillating one from the sub-leading instanton and after $N$ such periods we reach a critical axion value. In the alignment scenario this is where the quadratic approximation breaks down and the periodic nature of the system kicks in to censure the excursion distance, in the monodromy setting at the same axion value instead the strong backreaction kicks in and serves as the cutoff mechanism. In both cases although there are $N$ oscillations before the cutoff mechanism the oscillation period scales as $\frac{1}{N}$ thereby ensuring a cancellation in the proper excursion length.

\subsubsection{Calabi-Yau Models}
\label{sec:RRaxionCY}

So far we have restricted to the simplest setting where we have just a single representative from each moduli sector. We are now interested in studying how the results obtained are modified when we consider more involved CY compactifications with multiple moduli. We will consider still a simplified model where we take only a single representative variation in the Kahler moduli sector but will consider a more complicated complex-structure moduli sector. Let us consider the following setup
\bea
K &=& - \log s - 2\log \sqrt{u_1}\left(u_2-\frac23 u_1\right) - 3\log t \;, \nn \\
W &=& e_0 + i h_0 S - i h_1 U_1 - i h_2 U_2+ i e_1 T - q T^2 + \frac{i}{6} m T^3 \;. \label{cyex1}
\eea
The complex-structure moduli Kahler potential is the mirror of the $P_{[1,1,2,2,6]}$ CY studied in \cite{Candelas:1993dm}, see \cite{Grimm:2004ua,Palti:2008mg} for details of the mirror map for type IIA orientifolds. The massive axion combination we are interested in displacing is 
\be
\rho = e_0 - h_0 s + h_1 \nu_1 + h_2 \nu_2 \;.	
\ee
Let us look at its proper length (\ref{defphprsim}) 
\bea
\Delta \phi &=&  \int_{\rho_i}^{\rho_f} \sqrt{\frac{3}{2}}\left[6 h_0^2 s^2 + 6 h_1^2 u_1^2 + 8 h_1 h_2 u_1^2 + 
 h_2^2 \left(4 u_1^2 - 4 u_1 u_2 + 3 u_2^2\right)  \right]^{-\half}  d \rho \;, \nonumber \\
&=& \int_{\rho_i}^{\rho_f} \sqrt{\frac{3}{2}}\left[6 \tilde{s}^2 + \left(6  + 8 r + 4 r^2\right) \tilde{u}_1^2 - 4 r \tilde{u}_1 \tilde{u}_2 + 3 \tilde{u}_2^2 \right]^{-\half}  d \rho \;. \label{depex1}
\eea
Here we defined $\tilde{s} = h_0 s$, $\tilde{u}_i = - h_i u_i$, $r = \frac{h_2}{h_1}$. An important qualitative difference from the simpler one-modulus expression (\ref{delphigen}) is that there appears a flux parameter $r$ explicitly. This parameter is dimensionless under the symmetry (\ref{ssym}) and therefore may appear arbitrarily in the evaluation of the integral for $\Delta \phi$. At the same time it is also not possible to write the potential with only a one-parameter explicit dependence as in (\ref{1parpot}) and instead we have a two-parameter system $\tilde{V}\left(\tilde{s},\tilde{u}_1,\tilde{u}_2,\tilde{t},v',\rho',f,r\right)$. In order to see how the new parameter $r$ may affect the field distance we therefore need to study the moduli stabilisation equations. Proceeding as before we have the supersymmetric minimum
\be
\tilde{s}_0 = \frac{3 + 2 r}{3} \left(\tilde{u}_{1}\right)_0= \frac{3 + 2 r}{6\left(1 + r\right)} \left(\tilde{u}_2\right)_0=\frac{\tilde{t}_0^3}{15} = \frac29 \sqrt{\frac{10}{3}} f^{\frac32} \;,\; \rho_0 = \frac23 \tilde{q} \left(3f +2 \tilde{q}^2 \right)  \;,\; \tilde{v}_0 = -2\tilde{q} \;. \label{susymodex1}
\ee
We define $\rho'$ and $v'$ as in (\ref{rpvp}). Turning to solving the stabilisation equations (\ref{minpot}) we find a again a factorisation structure however now with four branches which determine both $\tilde{u}_1$ and $\tilde{u}_2$ in terms of $\tilde{s}$ and $\tilde{t}$ as
\bea
\mathrm{Branch\;1}:\; \tilde{u}_1 &=& \frac{3 \tilde{s}}{3 + 2r} \;,\;\; \tilde{u}_2 = \frac{6\left(1+r\right) \tilde{s}}{3 + 2r}   \;,\nonumber \\
\mathrm{Branch\;2}:\; \tilde{u}_1 &=& \frac{3 \tilde{s}}{3 + 2r} \;,\;\; \tilde{u}_2 = \frac13 \left(-\frac{6\left(3+r\right) \tilde{s}}{3 + 2r}+ \tilde{t}^3\right)   \;,\nonumber \\
\mathrm{Branch\;3}:\; \tilde{u}_1 &=& \frac{-6\tilde{s}+\tilde{t}^3}{2\left(3 + 2r\right)} \;,\;\; \tilde{u}_2 = \frac{6\left(3+r\right) \tilde{s}+ r \tilde{t}^3}{9 + 6r}  \;,\nonumber \\
\mathrm{Branch\;4}:\; \tilde{u}_1 &=& \frac{-6\tilde{s}+\tilde{t}^3}{2\left(3 + 2r\right)} \;,\;\; \tilde{u}_2 = -\frac{\left(1+r\right)\left(6\tilde{s}- \tilde{t}^3\right)}{3 + 2r}  \;. \label{branchescyex1}
\eea
Once we impose lying on any of the branches the remaining three equations in (\ref{minpot}) depend only on $\tilde{s}$, $\tilde{t}$, $v'$, $\rho'$ and $f$. Importantly the dependence on $r$ drops out which means that the solutions for $\tilde{s}$ and $\tilde{t}$ only depend on $f$ and $\rho'$. Further, once we impose a branch from (\ref{branchescyex1}) we find that the dependence on $r$ also drops out from the proper field length (\ref{depex1}). For example for branch 1 we have
\be
\Delta \phi = \int_{0}^{\rho'_{\mathrm{crit}}} \frac{1}{4\tilde{s}} \;d \rho' \;, \label{depex1br1}
\ee
which is the same expression as branch 1 for the one modulus case. After imposing branch 1 of (\ref{branchescyex1}), the equations determining $\tilde{s}$ are also the same as the one modulus case which means that the result is identical to (\ref{dpbr11m}). Similarly branch 4 of (\ref{branchescyex1}) reproduces the one modulus branch 2 case. The other new branches for the two modulus case give different results but the key property that the integrand in $\Delta \phi$ depends on only one flux parameter $f$ guarantees by the previous reasoning that the result is flux independent. In summary, we find that for this example of a more complicated CY setting with multiple complex-structure moduli there is additional structure in the relations between the $u_i$ and $s$ and an additional flux parameter in the potential, but in the proper field length this parameter drops out and the results is qualitatively and for some cases quantitatively the same as the one modulus case studied in detail in the previous subsection.

Let us present a different example CY. The complex-structure moduli Kahler potential is the mirror of the $P_{[1,1,1,6,9]}$ CY studied in \cite{Candelas:1994hw}. 
\be
K = - \log s - 2\log \left(u_1^{\frac32}-u_2^{\frac32}\right) - 3\log t \;.
\ee
The proper path length is given by
\be
\Delta \phi = \int_{\rho_i}^{\rho_f} \sqrt{\frac{3}{4}}\left[3 h_0^2 s^2 + 6 h_1 h_2 u_1 u_2 + h_2^2 \sqrt{u_2} (2 u_1^{\frac32} + u_2^{\frac32}) + 
 h_1^2 (u_1^2 + 2 \sqrt{u_1} u_2^{\frac32}) \right]^{-\half}  d \rho \;. \label{depex2}
\ee
Let us define 
\be
\tilde{s} = h_0 s \;, \tilde{T} = T m^{\frac13} \;, \tilde{e}_1 = e_1 m^{-\frac13} \;, \tilde{q} = q m^{-\frac23} \;,
\ee
and $f$ as in (\ref{fdef}). There is a supersymmetric vacuum at
\be
\tilde{s}_0 = -\frac{h_1^3+h_2^3}{3h_2^2} \left(u_2\right)_0= -\frac{h_1^3+h_2^3}{3h_1^2} \left(u_1\right)_0= \frac{\tilde{t}_0^3}{15} = \frac29 \sqrt{\frac{10}{3}} f^{\frac32} \;,\; \rho_0 = \frac23 \tilde{q} \left(3f +2 \tilde{q}^2 \right)  \;,\; \tilde{v}_0 = -2\tilde{q} \;. \label{susymodex2}
\ee
Note that the physical domain of fluxes is at $h_0>0$, $h_1<0$, $h_2>0$, and $\left|h_1\right|>\left|h_2\right|$. As before we shift the axions by their supersymmetric values (\ref{rpvp}) to define $\rho'$ and $v'$, and also define $r = \frac{h_2}{h_1}$. Turning now to the solutions to (\ref{minpot}) we find two physical branches
\bea
\mathrm{Branch\;1}:\; u_1 &=& -\frac{3 \tilde{s}}{h_1\left(1 + r^3\right)} \;,\;\; u_2 = -\frac{3 r^2\tilde{s}}{h_1\left(1 + r^3\right)}    \;,\nonumber \\
\mathrm{Branch\;2}:\; u_1 &=& \frac{6 \tilde{s}-\tilde{t}^3}{2h_1\left(1 + r^3\right)} \;,\;\; u_2 =  \frac{r^2\left(6 \tilde{s}-\tilde{t}^3\right)}{2h_1\left(1 + r^3\right)}  \;.
\eea
Restricting to these branches leads to remaining equations which are independent of the $h_i$ and which exactly match the corresponding equations for the one modulus case in the respective branches. Therefore the solutions for $\tilde{s}$ are the same again. Further, evaluating the integrand in (\ref{depex2}) over the branches leads to a cancellation of the $h_i$ fluxes leading again to a flux independent result as in the one modulus case. 

We have found the same behaviour also for other examples for CY Kahler potentials, and given that the cancellation of the fluxes in the final result appears to be very intricate it is reasonable to expect that there is an underlying reason or symmetry behind this which holds for any CY. However we were unable to prove this.

\subsubsection{Twisted Torus Models}
\label{sec:RRaxiontt}

In section \ref{sec:RRaxionCY} we studied a modification of the simplest model of section \ref{sec:RRaxionsingle} through adding structure to the Kahler potential. In particular this added another independent flux parameter making the system a two parameter one. In this section we modify the simplest model through a modification of the superpotential. This will lead again to a two flux parameter system.  We consider a compactification of type IIA string theory on a twisted torus.\footnote{More generally these can be considered as compactifications on a manifold with $SU(3)$-structure as in \cite{House:2005yc,Caviezel:2008ik}. In particular coset spaces are very tractable cases with few moduli.} The details of the compactification are given in section \ref{sec:iiatt}. We will study a simpler version of this model where we restrict the complex structure and Kahler moduli to a single variation, ie. take one representative in each class. The key addition to the models of section \ref{sec:RRaxionsingle} is an additional superpotential interaction which couples the dilaton and complex-structure moduli to the Kahler moduli. Due to this interaction both the RR axions gain a mass for generic fluxes. We will therefore restrict the fluxes further such that only one combination of the RR axions gains a (perturbative) mass while the other remains massless. This then singles out the massive direction as a monodromy axion along which we will displace and study the backreacted field excursion distance. The system we study is 
\be
K = - \log s - 3\log u - 3\log t \;, \;\;\; W = e_0 + i a l  S - i b l U + i e_1 T - q T^2 + \frac{i}{6} m T^3 + a S T -b T U \;. \label{wttbr}
\ee
The two new interactions and flux parameters are $a$ and $b$. We have restricted the NS flux $h_i$ to be of the form $h_0=a l$ and $h_1=b l$ with $l$ some free parameter which is only constrained by quantisation of the NS flux. This ensures that only one combination of the RR axions gains a mass. We can now perform some field redefinitions to make the system simpler. First we shift the axions and fluxes to set $e_0=q=0$ through
\bea
v &=& v' -\frac{2 q}{m} \;,\;\; \sigma = \sigma' + \frac{3 e_0 m^2 + 6 e_1 m q + 8 q^3}{3 a m \left(l m - 2 q\right)}\;, \nn \\
l' &=& l  - \frac{2 q}{m} \;,\;\; e_1' = \frac{3 e_0 m^2 + 3 e_1 l m^2 + 6 l m q^2 - 4 q^3}{3 l m^2 - 6 m q} \;.
\eea
We then define the rescaled fields
\be
T' = \frac{\tilde{T}}{m^{\frac13}} \;,\;\;  U = -\frac{ \tilde{l}m^{\frac13} \tilde{U}}{b} \;,\;\;  S' = \frac{\tilde{l}m^{\frac13} \tilde{S}}{ a}  \;,\;\; e_1' = \tilde{e}_1 m^{\frac13} \;,\;\; l' =\frac{1}{ \tilde{l} m^{\frac13}} \;.
\ee
The resulting superpotential and Kahler potential (up to an unimportant constant shift) then take the form
\be
K = - \log \tilde{s} - 3\log \tilde{u} - 3\log \tilde{t} \;, \;\;\; W =  i \left(\tilde{S} +  \tilde{U}\right) + i \tilde{e}_1 \tilde{T}  + \frac{i}{6} \tilde{T}^3 + \tilde{l} \tilde{T} \left(\tilde{S}+\tilde{U} \right)  \;. \label{wttres}
\ee
This makes it manifest that there are only two flux parameters in the system and that only one combination of RR axions gains a mass which we denote as $\rho= \tilde{\sigma} + \tilde{\nu}$. In the form (\ref{wttres}) we see that we recover the torus limit as studied in section \ref{sec:RRaxionsingle} by taking $\tilde{l}\rightarrow 0$.\footnote{Note that in terms of the original fluxes this is taking $a$,$b \rightarrow 0$, $l \rightarrow \infty$ with $a l$ and $b l$ finite.}

The system has a supersymmetric vacuum. In the limit $\tilde{l} \rightarrow 0$ this is the solution given in (\ref{susymod}). For non-vanishing $\tilde{l}$ it is given by \cite{Camara:2005dc}
\bea
\tilde{s}_0 = \frac{1}{3} \tilde{u}_0 = \frac{\tilde{v}_0 \tilde{t}_0}{\tilde{l}} \;,\;\; \tilde{t}_0^2 = \frac{15}{\tilde{l}} \tilde{v}_0\left(1+\tilde{v}_0\tilde{l} \right) \;,\;\; \rho_0 = - \frac{9 \tilde{v}_0+8\tilde{v}_0^2 \tilde{l} + 2 \tilde{e}_1 \tilde{l}}{2 \tilde{l}^2} \;,
\eea
and $\tilde{v}_0$ is given by the cubic equation
\be
160 \tilde{v}_0^ 3 \tilde{l}^2 + 186 \tilde{v}_0^2 \tilde{l} +27 \tilde{v}_0+6\tilde{e}_1 \tilde{l}=0 \;.
\ee
As before, it is useful to introduce the shifted axion field
\be
\rho' = \rho - \rho_0 \;.
\ee

\begin{figure}[tbp]
		\centering
		\includegraphics[width=0.7\textwidth]{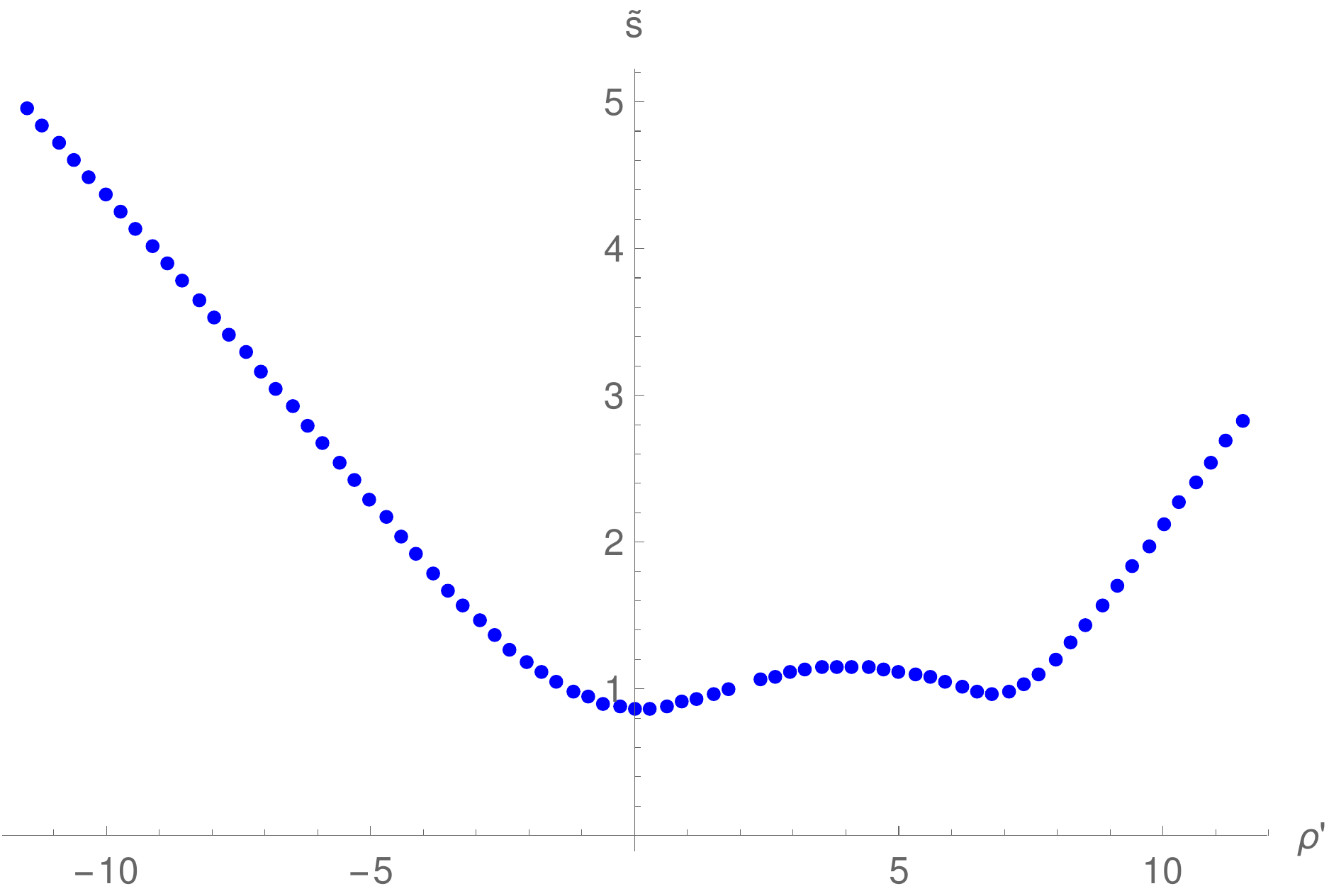}
		\caption{Plot showing $\tilde{s}$ as a function of $\rho'$ for the case of a twisted torus compactification with fluxes $\tilde{e}_1=-6$ and $\tilde{l}=1$.}
	\label{fig:stt}
\end{figure}

Turning to the scalar potential turning point equations (\ref{minpot}), we have two branches of solutions
\bea
\mathrm{Branch\;1}:\; \tilde{u} &=& 3 \tilde{s} \;, \nonumber \\
\mathrm{Branch\;2}:\; \tilde{u} &=& \frac{-6 \tilde{s} - 12 \tilde{v} \tilde{l} \tilde{s} - 6 \tilde{v}^2 \tilde{l}^2 \tilde{s} - 2 \tilde{l}^2 \tilde{s} \tilde{t}^2 + \tilde{t}^3}{2 \left(1 + 2 \tilde{v} \tilde{l} + \tilde{v}^2 \tilde{l}^2 - \tilde{l}^2 \tilde{t}^2\right)}  \;. \label{branchestt}
\eea
Restricting to branch 1, for negative $\tilde{e}_1$ the potential has three turning points at the supersymmetric value for $\rho'=0$, only one of which is the supersymmetric minimum.\footnote{We have also studied branch 2 in (\ref{branchestt}) and find that it behaves similarly to branch 2 in the model of section \ref{sec:RRaxionsingle}, which is shown in figure \ref{fig:second branch}, but with the behaviour of $\tilde{t}$ and $\tilde{u}$ exchanged.} We will restrict to studying excursions away from this minimum. The value of $\tilde{s}$ as a function of $\rho'$ is shown in figure \ref{fig:stt}. Contrary to the previous case, $\tilde{s}$ is not an even function of $\rho'$ and exhibits a second minimum. Studying this for different flux values we find that the second minimum is at positive values of $\rho'$ for positive values of $\tilde{l}$ and vice versa. We also find that outside the region between the two minima $\tilde{s}$ enters quite quickly a linear scaling regime asymptoting to $\tilde{s} = \alpha \rho'$ as in (\ref{amsolsim}). Inside the region $\tilde{s}$ remains approximately constant. We find that a good fit for the length of the approximately constant region is $\Delta\rho' \simeq 2\left(-\frac{\tilde{e}_1}{\tilde{l}^2}\right)^{\frac{3}{4}}$. We can thus approximate the proper field excursion length by taking $\tilde{s}$ to be approximately constant along $\Delta \rho'$. Therefore the value of $\tilde{s}$ is approximately its value at the supersymmetric minimum. This can be easily solved for analytically and we find that indeed $\tilde{s}_0 \simeq \left(-\frac{\tilde{e}_1}{\tilde{l}^2}\right)^{\frac{3}{4}}$. The result is therefore that to a good approximation $\Delta \phi$ is flux independent. As a measure of this we scanned over flux ranges $-100 \leq \tilde{e}_1 \leq -3$ and $1 \leq \tilde{l} \leq 100$ finding $2 \leq \Delta \phi\leq 3.5$. Such a small variation over such a large variation in $\Delta \rho'$ presents good evidence that $\Delta \phi$ is flux independent also for this setting.\footnote{Note that the fact that it is possible to reach values such as $\Delta \phi \simeq 3.5$ does not imply super-Planckian excursions since these are approximate values and are not reliable up to order one factors.} 

\subsection{Neveu-Schwarz axions}
\label{sec:NSaxion}

In the previous section we studied the displacement of the massive RR axions combination from the minimum of the potential. In this section we consider a different direction in field space away from the minima where we displace a massive combination of Neveu-Schwarz (NS) axions. Unlike the RR axions, where there is only a single combination which is perturbatively massive, generically all the NS axions will gain a mass from the fluxes in the compactification. There is therefore no distinguished direction in the NS axions field space to displace along with the exception of the case where all the moduli and axions are set to equal values and the displacement is then along this universal value. In this section we will study this case which is also the simplest setting. We leave a more complete study of various more complicated directions in a richer multiple NS axions setting for future work.

The theory we study is therefore the one in (\ref{simgen}), but we now impose 
\begin{equation}\label{minpotNS}
\partial_{t}V=\partial_{U}V=\partial_S V = 0 \;,
\end{equation}
while keeping the NS axion $v$ free to displace from the minimum. A large portion of the analysis of the previous section follows through for this case also. In particular the minima of the potential are the same, the factorisation into two branches in (\ref{branches}) continues to hold, the scaling symmetry (\ref{ssym}) still holds, and we can still write the potential as a one-parameter model (\ref{1parpot}).

Unlike the case for RR axions, both the branches in (\ref{branches}) support physical solutions for $f=0$ which read
\begin{align}\label{NSf=0}
\mathrm{Branch\;1}\;&:&\;\tilde{s} = 0.36 v'^3 \;,\;\; \tilde{u} = 1.07 v'^3 \;,\;\; \tilde{t} = 1.57 v' \;,\;\; \rho' = -0.17 v'^3 \;, \nonumber \\
\mathrm{Branch\;2}\;&:&\;\tilde{s} = 0.26 v'^3 \;,\;\; \tilde{u} = 1.19 v'^3 \;,\;\; \tilde{t} = 1.58 v' \;,\;\; \rho' = -0.17 v'^3 \;.
\end{align}
As before, these do not flow to a physical minimum due to the restriction $f=0$. However, as expected, they show the same logarithmic behaviour of the proper field distance in the axion $v'$ distance. 

\begin{figure}[tbp]
	\centering
	\begin{subfigure}[b]{0.5\textwidth}
		\centering
		\includegraphics[width=\textwidth]{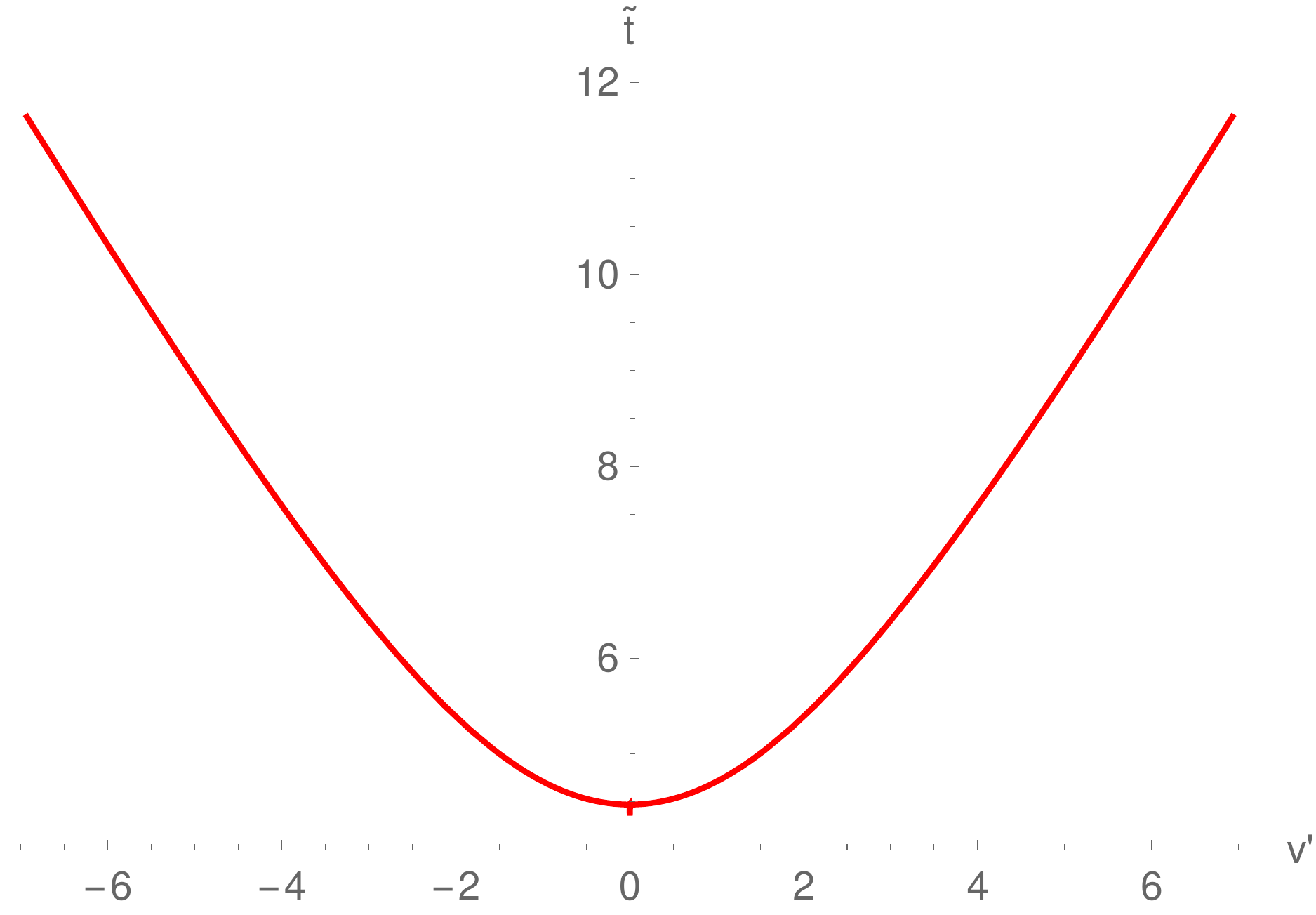}
		\caption{Branch 1}
	\end{subfigure}~
	\begin{subfigure}[b]{0.5\textwidth}
		\centering
		\includegraphics[width=\textwidth]{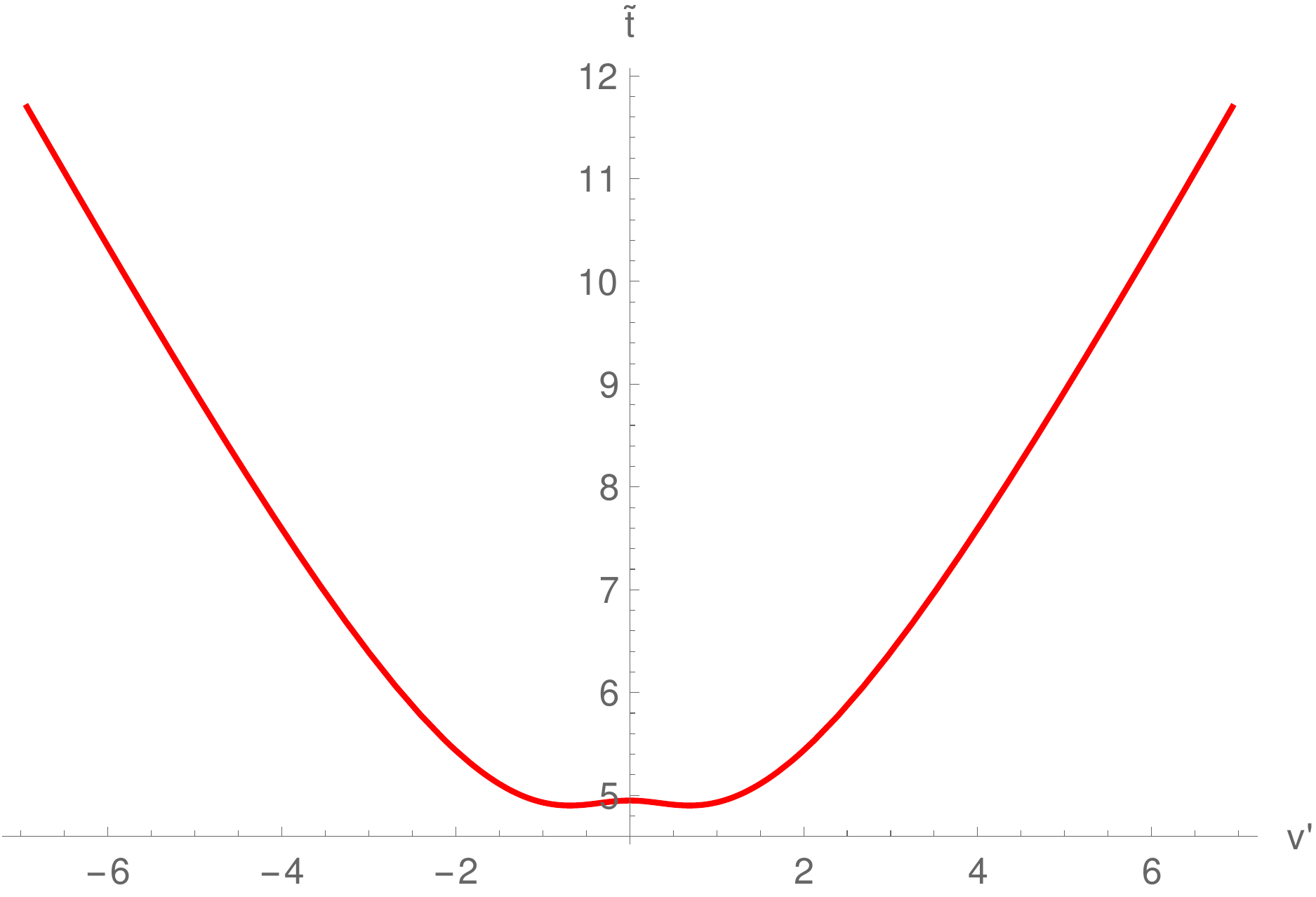}
		\caption{Branch 2}
	\end{subfigure}
	\caption{Plots showing the moduli $\tilde{t}$ as a function of $v' = v- v_0$ for displacement of $v'$ along both branches (\ref{branches}). They are given for flux value $f=6$. The range is chosen such that it is possible to see the asymptotic linear behaviour is reached after $v'$ reaches its critical value \eqref{NScritVal}.}
	\label{fig:NSaxions}
\end{figure}

Once we turn on $f \neq 0$ the solutions become more complicated, but are still simpler than the RR axion case such that we can obtain an analytic expression for them. We find that $\tilde{t}$ as a function of $v'$ is given by the solution of the following equations, depending on the branch:
\begin{align}
\mathrm{Branch\;1}\;&:\;
25 v'^6+35 \tilde{t}^2 v'^4+8 f (33 \tilde{t}^4 v'^2)+8 f^2 (33 \tilde{t}^4+35 \tilde{t}^2 v'^2+75 v'^4)\nonumber\\
&\;+25 v'^8+70 \tilde{t}^2 v'^6+115 \tilde{t}^4 v'^4+6 \tilde{t}^6 v'^2+800 f^3 v'^2-27 \tilde{t}^8+400 f^4=0\nonumber\\
\mathrm{Branch\;2}\;&:\;25 v'^8+10 (20 f+7 \tilde{t}^2) v'^6+(600 f^2+280 f \tilde{t}^2+43 \tilde{t}^4) v'^4\nonumber\\
&\;+2 (400 f^3+140 f^2 \tilde{t}^2-12 f \tilde{t}^4-15 \tilde{t}^6) v'^2+8 f^2 (50 f^2-3 \tilde{t}^4)=0
\end{align}
The relevant solution for each branch is a known complicated function of $v'$ and $f$ and will not be explicitly shown. Instead we show their behaviour in figure \ref{fig:NSaxions}. One can see that for large values of $v'$, we get back to the behaviour of equation \eqref{NSf=0}. This happens after a critical value
\begin{equation}\label{NScritVal}
	v'_\textrm{crit}\simeq \sqrt{2f}\;.
\end{equation}

Again, in complete analogy with the RR axions, we can find the proper distance traveled by the NS axions up to that critical value. Because of the symmetry \eqref{ssym}, we expect it to be independent of flux values. This is indeed the case, as one gets:
\begin{equation}
	\Delta\phi=\frac{\sqrt{3}}{2}\int_0^{v'_\textrm{crit}}\frac{dv'}{\tilde{t}}\simeq
	\begin{cases}
		0.57\qquad&\textrm{Branch 1}\\
		0.55\qquad&\textrm{Branch 2} 
	\end{cases}
\;.
\end{equation}
As before, any further excursion will add only logarithmic corrections to those values.

With regards to the stability analysis of the potential. Moving away from the turning points at $v'=\rho'=0$ discussed in section \ref{sec:RRaxionsingle}, an analysis of the Hessian matrix with respect to the other directions shows that despite one of the eigenvalues picking up a negative sign for some values of $v'$, it always lies above the BF bound. This stability holds for both branches, as well as both signs of $f$. Moreover plugging back the fields satisfying \eqref{minpotNS} in the potential such that it only depends on $v'$ and the flux numbers, this expression has a local minimum at $v'=0$ for both signs of $f$ in branch 2. For branch 1, it has a global minimum for positive $f$ and local maximum for negative $f$.

Finally we also studied excursions along the NS axions in the twisted torus setting of section \ref{sec:RRaxiontt}. We find a similar structure to the RR axion case with a range over which $\tilde{t}$ is approximately constant bounded by linear scaling regimes. We find that a good fit for the length of the approximately constant region is $\Delta v' \simeq 3 \left( -\frac{\tilde{e}_1}{\tilde{l}}\right)^{\frac13}$. This seems to be a good fit to the supersymmetric value $\tilde{t}_0$. Scanning over the flux ranges $-100 \leq \tilde{e}_1 \leq -3$ and $1 \leq \tilde{l} \leq 100$ we find $1.8 \leq \Delta \phi\leq 2.0$, which presents good evidence that it is flux independent.

\section{Axion alignment and backreaction}
\label{sec:axionalign}

In the previous section we studied the backreaction of the axion vev on the axion field-space metric in an axion monodromy scenario. In this section we consider an axion alignment scenario, as in \cite{Kim:2004rp} but where the alignment is induced through fluxes \cite{Hebecker:2015rya,Palti:2015xra}. From a field theory approach, where we can choose the axion field space metric freely, such a setup leads to an effective super-Planckian decay constant for the axions combination which remains massless after the mixing. In \cite{Palti:2015xra} an analysis of this scenario in type IIA string theory concluded that, for all the classes of CY spaces studied, the large parameter $N$ responsible for the mixing backreacts on the axion periods in such a way as to censure this super-Planckian enhancement. It was found that this occurs in two qualitatively different ways depending on the structure of the Kahler potential for the geometric moduli. For `toroidal-like' Kahler potentials    it was shown that there is a scaling of the fundamental axion decay constants in such a way as to cancel any enhancement of the effective axion decay constant. For other types of Kahler potentials it was shown that this cancellation does not occur however there was always a sub-Planckian period instanton which dominated in magnitude over the enhanced period instanton. The analysis performed in \cite{Palti:2015xra} was primarily of two fundamental axions mixing to form an effective massless axion combination. Some analysis of three axions mixing was performed and there it was shown that while for type IIA CY compactifications this does not lead to an effective super-Planckian decay constant, in principle if the superpotential was modified such that there was a new mass term to fix certain axion combinations then the possibility of super-Planckian enhacement was not ruled out. It was suggested that such an additional superpotential contribution could arise in type IIB compactifications on a CY or in type IIA compactifications on a twisted torus (or more generally a manifold with torsion). In this section we will therefore analyse such a setup with more than two axions (more precisely four axions) mixing and additional superpotential contributions for type IIA compactifications on a twisted torus. 

\subsection{Three axion alignment toy model}
\label{sec:3atoy}

We have presented some motivation for considering mixing between more than two fundamental axions. Before proceeding to a detailed analysis of a full model let us present a toy model which captures the effect that we are interested in. First recall the setting of two axions mixing in a toroidal-like model. We have a setup with two complex scalar fields, $U_i=u_i+i\nu_i$ with $i=1,2$, whose imaginary components are axions in that they do not appear in the Kahler potential. We would like to induce alignment between these axions by lifting one combination of them through a superpotential $W = W\left( q_1 U_1 - q_2 U_2\right)$. 
Consider a supersymmetric minimum of this system, the F-terms then imply
\be
\frac{K_{u_1}}{K_{u_2}} = \frac{q_1}{q_2} \;, \label{modfterms}
\ee
where $K_{u_i} = \frac{\partial K}{\partial u_i}$. 
We take the toy toroidal-like Kahler potential 
\be
K = -\ln u_1 -\ln u_2 \;.
\ee
Then if we fix the moduli supersymmetrically we can consider the F-terms ratio giving
\be
\frac{K_{u_1}}{K_{u_2}} =\frac{f_{u_1}}{f_{u_2}} =\frac{u_2}{u_1} = \frac{q_1}{q_2} \;.
\ee
Now if we take $q_1 \gg q_2$ we have two options, either to keep $u_2$ and therefore $f_{u_2}$ constant and take $u_1 \ll 1$ or to keep $u_1 > 1$ and take $u_2 \gg 1$ and therefore $f_{u_2} \ll 1$. The former option leads to a region in moduli space which is not under control. The latter option leads to the effect that the axion decay constants scale with the flux precisely in such a way to cancel the possible enhancement of an effective decay constant since the latter scales as $f_{\mathrm{eff,1}} \sim \frac{q_1f_{u_2}}{q_2}$ \cite{Palti:2015xra}.

Consider now a setting where we have three axions and two superpotential contributions
\be
K = -\ln u_1 -\ln u_2 - \ln u_3 \;,\;\; W = W_1\left(q_1 U_1 + q_2 U_2 + q_3 U_3\right) + W_2\left(\bar{q}_1 U_1 + \bar{q}_2 U_2 \right) \;.
\ee
Now it is important that $W_1$ and $W_2$ are not just linearly added so that they fix two combinations of axions. To illustrate the result we can take $q_1=-q_2=q$ and $\bar{q}_1=\bar{q}_2=1$ in which case we find from a similar analysis of the F-term ratios
\be
\frac{K_{u_2}-K_{u_1}}{K_{u_3}} = \frac{f_{u_2}-f_{u_1}}{f_{u_3}} = \frac{2q}{q_3} \;.
\ee 
Now taking $q_3$ large does not imply a scaling of the axion decay constants to cancel it, indeed it is compatible with having $f_{u_1} \sim f_{u_2}$ and free in magnitude. The conclusion is that for more than two fundamental axions mixing the single constraint on the moduli values coming from the ratio of the F-terms is not sufficient to show that no super-Planckian enhancement is possible. Therefore if such a constraint emerges it must do so from a more detailed analysis of moduli stabilisation. 

\subsection{Axion alignment on a twisted torus}
\label{sec:iiatt}

In the previous subsection we showed that for more than two axions mixing any obstruction from flux backreaction to an enhanced effective decay constant must manifest through a more detailed analysis of moduli stabilisation, in this section we perform such an analysis for a compactification of type IIA string theory on a twisted torus. We will consider the setup studied in \cite{Camara:2005dc}. The manifold has intrinsic torsion which means it has a set of non-closed 1-forms
\be
d \eta^{P} = - \frac12 w_{MN}^P \eta^{M} \wedge \eta^{N} \;, \label{de}
\ee
where $M=1,..,6$ are the six toroidal directions.
The structure constants $w_{MN}^P$ have the following properties
\be
w_{MN}^P = w_{[MN]}^P \;, \;\; w_{PN}^P = 0 \;, \;\; w_{[MN}^P w^S_{L]P} = 0 \;. \label{wprott}
\ee
The last equation in (\ref{wprott}) comes from the nilpotency of the exterior derivative applied to (\ref{de}). It is called the Jacobi identity and plays an analogous role to tadpole constraints.
It is convenient to introduce labels for the non-vanishing components of the torsion
\bea 
\bmat{c} a_1 \\ a_2 \\
a_3 \emat = \bmat{c} \om^1_{56} \\ \om^2_{64} \\ \om^3_{45} \emat
\quad , \quad \bmat{ccc} b_{11} & b_{12} & b_{13} \\ b_{21} & b_{22}
& b_{23} \\ b_{31} & b_{32} & b_{33} \emat = \bmat{ccc} \! \! \!
-\om^1_{23} & \, \om^4_{53} & \, \om^4_{26} \\ \, \om^5_{34} & \! \!
\! -\om^2_{31} & \, \om^5_{61} \\ \, \om^6_{42} & \, \om^6_{15} & \!
\! \! -\om^3_{12} \emat \ .
\label{abmatrix}
\eea 
The Jacobi identities imply the twelve constraints  
\bea 
b_{ij}
a_j + b_{jj} a_i & = & 0 \quad , \quad i \not= j \nonumber \;,\\[0.2cm]
b_{ik} b_{kj} + b_{kk} b_{ij} & = & 0 \quad , \quad i \not= j \not= k
\;,
\label{jacb} 
\eea 
with indices $i,j=\left\{1,2,3\right\}$. The resulting superpotential takes the form \cite{Camara:2005dc}
\bea 
W & = & e_0
+ ih_0 S + \sum_{i=1}^3 \left[\left(ie_i - a_i S - b_{ii}U_i -\sum_{j\not= i}
b_{ij}U_j\right)T_i - i h_iU_i\right] \nonumber \\[0.2cm] & - & q_1 T_2 T_3 -q_2
T_1 T_3 -q_3 T_1 T_2 + i m T_1 T_2 T_3 \;,
\label{wa}
\eea
and the Kahler potential is
\beq 
K= - \log s - \sum_{i=1}^3 \log u_i - \sum_{i=1}^3 \log t_i \;.
\label{kpot}
\eeq
Now we see that for general fluxes there are 4 complex-structure and dilaton axions $\left\{\sigma,\nu_i\right\}$ and 4 combinations which appear in the F-terms:
\bea
\psi_0 &=& -h_0 \sigma + h_1 \nu_1 + h_2 \nu_2 + h_3 \nu_3  \;, \nn \\
\psi_1 &=& a_1 \sigma + b_{11} \nu_1 + b_{12} \nu_2 + b_{13} \nu_3 \;, \nn \\
\psi_2 &=& a_2 \sigma + b_{21} \nu_1 + b_{22} \nu_2 + b_{23} \nu_3 \;, \nn \\
\psi_3 &=& a_3 \sigma +  b_{31} \nu_1 + b_{32} \nu_2 +b_{33} \nu_3  \;.
\eea
Note that the superpotential appears in the F-terms as
\bea
\mathrm{Re}(W) &\supset& \psi_0 + \psi_1 b_1 + \psi_2 b_2 + \psi_3 b_3 \;, \nn \\
\mathrm{Im}(W) &\supset& -\psi_1 t_1 - \psi_2 t_2 - \psi_3 t_3 \;.
\eea
We can write
\be
\left(\begin{array}{cccc}
-h_0 & h_1 & h_2 & h_3 \\ 
a_1 & b_{11} & b_{12} & b_{13} \\ 
a_2 & b_{21} & b_{22} & b_{23} \\ 
a_3 & b_{31} & b_{32} & b_{33} \\ 
\end{array}\right)
\left(\begin{array}{c}
\sigma \\ 
\nu_1 \\ 
\nu_2 \\ 
\nu_3 \\ 
\end{array}\right)
=
\left(\begin{array}{c}
\psi_0 \\ 
\psi_1 \\ 
\psi_2 \\ 
\psi_3 \\ 
\end{array}\right)
\;. \label{flmat}
\ee

We are interested in configurations which keep 1 unfixed axion direction, for which we would like to study the resulting effective decay constant. In \cite{Camara:2005dc} the solution for the Jacobi identity used was
\be
b_{ji}=b_i \;,\;\; b_{ii}=-b_i\;,\;\;a_i=a \;. \label{jacsol}
\ee
Adopting this solution the flux matrix in (\ref{flmat}) becomes\footnote{There are other solutions possible which lead directly to a massless axions without further restriction on the fluxes. For example $b_{ji}=b_i \;,\;\; b_{ii}=-b_i\;\;\;a_i=a \;,\;\; i=2,3 \;,\;\;a_1=b_{1i}=0$. However the resulting system is more difficult to solve and so we leave a study of such possibilities for future work.}
\be
\left(\begin{array}{cccc}
-h_0 & h_1 & h_2 & h_3 \\ 
a & -b_{1} & b_{2} & b_{3} \\ 
a & b_{1} & -b_{2} & b_{3} \\ 
a & b_{1} & b_{2} & -b_{3} \\ 
\end{array}\right)
\;. \label{ffluxmatrix}
\ee
To keep a massless axion we need to reduce the rank of the flux matrix. For general $h_i$ we can do this by setting two fluxes from the set $\left\{a,b_1,b_2,b_3\right\}$ to zero. However then the mixing arises between only two fundamental axions, ie. the eigenvector of the flux matrix with zero eigenvalue only has two non-vanishing components. Since we are interested in mixing more than two axions we take instead the restriction
\be
h_0=-3 l a  \;,\;\; h_i=l b_i  \;, \label{axc2}
\ee
with $l$ an arbitrary constant (up to quantisation requirements). This leaves a massless axion because then the first row in the flux matrix (\ref{ffluxmatrix}) is proportional to the sum of the last three. The remaining massless combination is 
\be
\psi = \psi_v \cdot \nu_v \equiv \left(b_1b_2b_3,-ab_2b_3,-ab_1b_3,-ab_1b_2 \right)\cdot\left(\sigma,\nu_1,\nu_2,\nu_3\right) \;.
\ee
It is therefore a mixture of all of the fundamental axions. The superpotential with this restriction takes the form
\bea
W & = & e_0 - 3 i l a S - i l b_1 U_1 - i l b_2 U_2 - i l b_3 U_3 \nn \\
&-& a S \left(T_1 + T_2 + T_3 \right) 
+ b_1 U_1 \left( T_1 - T_2 - T_3 \right)
+ b_2 U_2 \left( -T_1 + T_2 - T_3 \right)
+ b_3 U_3 \left( -T_1 - T_2 + T_3 \right) \nn \\
& + & i e_1 T_1 + i e_2 T_2 + i e_3 T_3 -q_1 T_2 T_3 -q_2 T_1 T_3 -q_3 T_1 T_2 + i m T_1 T_2 T_3 \ .
\label{waci2}
\eea
The perturbatively massless axion $\psi$ appears in four instantons associated to $s$ and the $u_i$. The effective axion decay constants for $\psi$ in the four instantons are given by
\be
\left(\begin{array}{c}
f_{\psi}^s \\ 
f_{\psi}^{u_1} \\ 
f_{\psi}^{u_2} \\ 
f_{\psi}^{u_3} \\ 
\end{array}\right)
  = f  \left(\begin{array}{c}
-a \\ 
b_1 \\ 
b_2 \\ 
b_3 \\ 
\end{array}\right)\;,  \label{eadcv}
\ee
with 
\be
f^2 = \frac{1}{a^2s^2} + \frac{1}{b_1^2u_1^2}+ \frac{1}{b_2^2u_2^2}+ \frac{1}{b_3^2u_3^2} \;. \label{fexp}
\ee
Now for constant values of the moduli we can induce a parameterically large effective axion decay constant by taking any of $\left\{a,b_1,b_2,b_3\right\}$ large. In the case of two fundamental axions mixing we could show an obstruction to this by using the vanishing F-term combination associated to the massless axion. Here this implies the relation
\be
\psi_v \cdot \left(K_{s},K_{u_1}, K_{u_2},K_{u_3}\right) = 0 \;,
\ee
which gives
\be
\frac{1}{a s}-\frac{1}{b_1 u_1}-\frac{1}{b_2 u_2}-\frac{1}{b_3u_3} = 0 \;.\label{frw}
\ee
If we restrict this to only two axions we obtain the familiar relation between ratios of decay constants and this leads to the usual cancellation \cite{Palti:2015xra} . However in this more general case this particular formulation of the obstruction is absent.

We therefore need to study the full moduli stabilisation scenario in more detail. Before doing so let us discuss the tadpole constraints which for general fluxes read
\bea
N_{D6} + \half\left(h_0 m + a_1 q_1 + a_2 q_2+ a_3 q_3 \right) = 16 \;,\nn \\
N_{D6} + \half\left(h_i m - b_{1i} q_1 - b_{2i} q_2- b_{3i} q_3 \right) = 0 \;. \label{tadpoles}
\eea
Imposing the constraints on the fluxes (\ref{jacsol}) and (\ref{axc2}) gives
\bea
N_{D6} + \half a \left(-3l m + q_1 + q_2+ q_3 \right) = 16 \;,\nn \\
N_{D6} + \half b_1 \left(l m + q_1 - q_2- q_3 \right) = 0 \;,\nn \\
N_{D6} + \half b_2 \left(l m - q_1 + q_2- q_3 \right) = 0 \;,\nn \\
N_{D6} + \half b_3 \left(l m - q_1 - q_2 + q_3 \right) = 0 \;. \label{tadpoles2}
\eea
The parameters which we need to dial up to enhance the axion decay constant are $a$ and the $b_i$. In the case of a CY compactification \cite{Palti:2015xra}  these were restricted by the tadpole constraints. Here they are unconstrained by tadpoles if we choose
\be
q_1=q_2=q_3\equiv q=l m \;. \label{freetad}
\ee
This shows that an obstruction to arbitrarily large axion alignment need not arise from the tadpole constraints. Apart from allowing for arbitrarily large alignment parameters the flux restriction (\ref{freetad}) is also special in terms of uplifting the four-dimensional vacuum to a ten-dimensional one. The tadpoles are topological charge cancellation conditions where in general flux contributions can cancel against localised sources. However they originate as the integration of the ten-dimensional Bianchi identity which is instead a local equation. The fluxes are taken to be smoothly distributed throughout certain cycles in the space and therefore cannot pointwise cancel against localised sources in the Bianchi identity. This usually leads to the approach of smearing the localised sources. This condition (\ref{freetad}) implies that localised sources and fluxes can cancel separately in the Bianchi identity leading to a better controlled ten-dimensional solution.

We now consider the most general solution for the F-terms of the superpotential (\ref{waci2}). Note that we use the notation (\ref{sfexpa}) for the superfield components. First we can solve for $s$ in terms of the $t_i$ and $v_i$ by considering
\be
\R \left( F_{U_1}+F_{U_2}+F_{U_3}-F_{T_1}-F_{T_2}-F_{T_3} \right) = 0 \;,
\ee
which gives
\be
a s=-\frac{2 \left(v_3 m t_1 t_2 + q_3 t_1 t_2 + v_2 m t_1 t_3 + q_2 t_1 t_3 + v_1 m t_2 t_3+ q_1 t_2t_3\right)}{
 t_1 + t_2 + t_3}\;.
\ee
Now by considering
\be
\R \left( F_S - F_{U_i} \right)=0 \;,
\ee
we can deduce
\be
b_1 u_1 = \frac{s \left(t_1+t_2+t_3\right)}{-t_1+t_2+t_3} \;,\;\;\;b_2 u_2 = \frac{s \left(t_1+t_2+t_3\right)}{t_1-t_2+t_3} \;,\;\;\; b_3 u_3 = \frac{s \left(t_1+t_2+t_3\right)}{t_1+t_2-t_3} \;. \label{modval}
\ee
Note that the precise locus where these expressions break down $t_1=t_2+t_3$ is not a viable solution since it implies $a=b_2=b_3=0$ which leaves us with flat directions. However in principle we may still approach this locus parametrically. This is important because the expression for $f$ can be written as
\be
f^2= \frac{4}{\left(as\right)^2} \frac{t_1^2+t_2^2+t_3^2}{\left(t_1+t_2+t_3\right)^2}= \frac{4}{\left(b_1 u_1\right)^2} \frac{t_1^2+t_2^2+t_3^2}{\left(-t_1+t_2+t_3\right)^2} \;.
\ee
This shows that is not possible to obtain an enhancement of the decay constant by taking $a$ large for any value of the $t_i$. However by taking $b_1$ large we could in principle obtain an enhancement if simultaneously we approach the limit $t_1 \rightarrow t_2+t_3$. 

Continuing with the solution analysis we consider the three combinations $\I \left(F_{S}-F_{U_i}\right)$, these take the form $A\left(t_i,v_i\right) F_i \left(t_i,v_i\right)$. It can be shown that imposing $A\left(t_i,v_i\right)=0$ leads to an unphysical solution $u_i =0$. While imposing $F_i \left(t_i,v_i\right)=0$ leads to 
\be
v_2 = \frac{-l t_1+l t_2+v_1t_2}{t_1}\;,\;\;v_3 = \frac{-l t_1+l t_3+v_1t_3}{t_1}\;.
\ee
We therefore arrive at a solution for $s$, $u_i$, $v_2$, $v_3$ in terms of $v_1$ and the $t_i$. 

Proceeding generally past this point is quite difficult. However recall that there was a special choice of fluxes (\ref{freetad}) which was required to be able to take the alignment $a$ and $b_i$ parameters large and unconstrained by the tadpoles. If we impose these restrictions then it can be shown that the F-term combinations $\R \left(F_{T_i}-F_{T_j}\right)$ lead to the constraint $t_1=t_2=t_3$. Using this in (\ref{modval}) then gives
\be
3as = b_i u_i \;,\;\; (\mathrm{no\;sum\;over\;i})\;. \label{scc}
\ee
For completeness we present the full solution for the geometric moduli here
\be
t = \frac{\sqrt{3}5^{\frac16}\left(e_0 + l\left(e_1+e_2+e_3+2 m l^2 \right) \right)^{\frac13}}{2^{\frac53}m^{\frac13}} \;, \;\;
s =\frac{2mt^2}{\sqrt{15}a} \;,\;\;
u_i = \frac{3 a s}{b_i}\;.
\ee

The important result is the expression (\ref{scc}) which is a significantly stronger constraint on the moduli values than (\ref{frw}). Indeed using (\ref{scc}) in the expression for the effective axion decay constants (\ref{eadcv}) leads to a cancellation of the fluxes in any enhancement. Therefore the effective decay constant remains sub-Planckian for any flux values. We therefore recover the same obstruction that we found in the two axion toroidal setting also in this qualitatively more involved setting. It is important to note that due to the complexity of the system we were only able to show this cancellation for the special flux choice (\ref{freetad}).

\section{Summary}
\label{sec:summary}

In this paper we studied axion monodromy scenarios in flux compactifications of type IIA string theory on CY manifolds and a twisted torus. In particular we calculated the effect that the backreaction of the axion vev has on the proper length in field space traversed by the axion. We found universal behaviour in all the settings studied where the backreaction is small up to some critical axion value. After this critical point the backreaction is strong and cuts off the proper field distance so that it increases at best logarithmically with the axion vev, with a proportionality factor between the proper field distance and the logarithm which is flux independent. The critical axion value can be made arbitrarily large through flux choices thereby allowing for arbitrarily large changes  in the axion vev, however the backreaction of the flux on the axion field space metric implies an exact cancellation of the fluxes in the proper distance traversed by the axion up to its critical value such that the result is flux independent and sub-Planckian. 

In more detail, we first studied the case where the monodromy axion is the single linear combination of RR axions which appears in the superpotential. Our starting point was the simplified model studied in \cite{Palti:2015xra} where the moduli are restricted to only one variation in each sector and some of the fluxes are turned off.   In this model the axion vev backreacts strongly and the proper field length is only logarithmic in the axion vev. We argued that this can be attributed to a scaling symmetry of the fields, and in particular using this symmetry showed that the logarithmic behaviour holds for any CY compactification, as long as the fluxes which were turned off remain so. Next we considered turning the remaining fluxes, which carried dimensions under the scaling symmetry and therefore served as order parameters for its breaking, back on. We showed that in this case, for the single variation simplified models, the logarithmic strong backreaction behaviour can be delayed arbitrarily far in the axion value by breaking the scaling symmetry arbitrarily strongly using the fluxes. However the same fluxes which shielded the moduli from the strong axion backreaction in turn backreacted themselves on the axion field space metric such that a precise cancellation occurred and the proper field distance up to the axion critical value was flux independent. We then generalised these results to the case of multiple complex-structure moduli and studied example CY Kahler potentials. We showed that in this case the system becomes more complicated and in particular depends on two combinations of flux parameters (rather than one parameter for the single modulus case). However the fluxes again canceled out in the proper field distance up to the critical value. Finally we studied compactifications on a twisted torus which in particular introduces new interactions in the superpotential. This setup can be reduced to a two flux parameters system. Solving for the backreacted moduli behaviour as a function of the axion vev numerically we showed that they take the form of an approximately constant region bounded by linear regimes for axion vev beyond some critical value. By scanning over flux parameters we showed that the proper field distance up to the critical axion value is again flux independent. After this we turned to excursion distances for NS axions and showed that they have the same qualitative behaviour as the RR axions described above.

We also studied an axion alignment model from compactifications of type IIA string theory a twisted torus where up to four fundamental axions mix. Neglecting the flux backreaction on moduli the resulting effective axion decay constant can be parametrically enhanced through flux choices. There is a particular choice of fluxes which allows the enhancement parameter to be arbitrarily large while remaining compatible with tadpole constraints. We calculated the flux backreaction on the moduli for such fluxes and find that the enhancement parameter cancels in the effective axion decay constant implying that it remains sub-Planckian.

The cancellation of the flux parameters in the result for the proper length in field space traversed by the axion in the axion monodromy scenarios and the cancellation of the fluxes in the effective decay constant in the axion alignment scenarios (as studied in more detail in \cite{Palti:2015xra}) are strikingly similar physics. 
Indeed we showed that the two can be naturally related by zooming into the origin of the large effective axion period in an axion alignment scenario where the potential looks approximately quadratic rather than sinusoidal and therefore is similar to an axion monodromy scenario. On top of the quadratic potential there are oscillations from the sub-leading instanton and after $N$ such periods we reach a critical axion value. In the alignment scenario this is where the quadratic approximation breaks down and the periodic nature of the system kicks in to censure the excursion distance, in the monodromy setting at the same axion value instead the strong backreaction kicks in and serves as the cutoff mechanism. In both cases although there are $N$ oscillations before the cutoff mechanism the oscillation period scales as $\frac{1}{N}$ thereby ensuring a cancellation in the proper excursion length. 

The results of this work uncover new mechanisms in string theory which censure super-Planckian axion excursions in axion monodromy and axion alignment scenarios. We did so in the specific type IIA setting because it is the simplest and best understood framework to calculate such effects since full moduli stabilisation can be realised by using only tree-level expressions for the Kahler potential and superpotential. However it is not unreasonable to expect that the mechanisms which operated in our settings could do so also in other string theory setups. In particular it would be interesting to study how the axion vev backreaction in axion monodromy scenarios affects its proper field distance in other string theory constructions and work towards a general understanding of this effect.

The fact that in all the cases studied we find that the growth rate of the proper field distance of the axion excursion beyond some critical point is at best logarithmic in the axion vev matches the general conjecture made within the swampland context of a maximum logarithmic growth for any field at field values asymptoting to infinity \cite{Ooguri:2006in}. The non-trivial cancellation of the fluxes in the proper field distance before the logarithmic behaviour kicks in lends some weight to a sharpened swampland conjecture. As well as conjecturing that the growth rate of the proper distance is at best logarithmic in a field for field values asymptoting to infinity we could conjecture that this logarithmic growth rate must set in at a sub-Planckian proper path distance.\footnote{There is a finite size transition region around the critical axion value between the small backreaction region and the linear scaling strong backreaction regime. This is the transition region to the logarithmic growth regime. The conjecture is that it begins at sub-Placnkian values and that the point where logarithmic growth is a better description than linear growth is also sub-Planckian.} The scenarios studied in this work present non-trivial tests of such a sharpened statement.

Let us discuss two cases which are relevant for the possibility of such a conjecture. First there is a simple case of fields where it can be understood through the WGC: moduli with axion superpartners. Using the WGC applied to axions implies a relation between the magnitude of the instanton, which is the modulus vev $u$, and the axion decay constant $f_a$ \cite{ArkaniHamed:2006dz,Brown:2015iha}
\be
S_{\mathrm{Inst}} f_a = u f_a \leq M_p \;.
\ee
This is typically used to bound the axion decay constant because if $f_a \geq M_p$ then $u < 1$ and therefore control over the instanton expansion is lost. However we can also use the same relation to bound the magnitude of $u$. Supersymmetry implies that $f_a=\sqrt{g_{uu}}$ and so is the measure on the modulus field space. We therefore have
\be
\sqrt{g_{uu}} u \leq M_p \;.
\ee
Therefore for $u>M_p$ we must have that $\sqrt{g_{uu}}$ decays at least with a power of $\frac{1}{u}$. This establishes that the proper field distance enters (at best) logarithmic growth at this point.

The second important case appears to present a counter example to at least a simple version of the conjecture. In  \cite{Palti:2015xra} an axion alignment model was found where one of the instantons did indeed develop a parametrically large axion decay constant.\footnote{This is the model in section 3.2 of \cite{Palti:2015xra} with $q^s \gg q^2 \gg q^1$ in which case the axion partner of the dilaton can be integrated out and the model reduces to the one of section 3.1.} However this instanton was found to always be sub-dominant to an instanton with an un-enhanced decay constant. This is consistent with the strong version of the WGC. While such a setup would not lead to a viable large field inflation model, it nonetheless appears to present an example where the axion displacement can be parametrically large while backreaction effects are exponentially small (coming from an instanton energy density competing against tree-level effects). Therefore either there are some inconsistencies in this alignment setup or the logarithmic behaviour conjecture must be refined appropriately, perhaps to account for the relative magnitude of the potential contributions for the field analogously to the refinement made in the strong WGC.

{\bf Acknowledgments}

We would like to thank Arthur Hebecker, Timo Weigand and Lukas Witkowski for useful and insightful discussions. In particular we thank Arthur Hebecker and Lukas Witkowski for collaboration during the initial stages of the project. 
EP would like to thank the Weizmann Institute for hospitality while this work was being completed. Our work is supported by the Heidelberg Graduate School for Fundamental Physics. 


\end{document}